\documentclass[11pt, a4paper]{article}
\usepackage[dvipsnames]{xcolor}
\usepackage[utf8]{inputenc}
\usepackage{url}
\usepackage{amsmath}
\usepackage{amsfonts}
\usepackage{amssymb} 
\usepackage{graphicx}
\usepackage{array}
\usepackage{lipsum}
\usepackage{float}
\usepackage{multirow}
\usepackage{hhline}
\usepackage{tabularx}
\usepackage{subfigure}
\usepackage{afterpage}
\usepackage{epstopdf}
\usepackage[utf8]{inputenc}

\usepackage[pdftitle={The Axion is Going Dark},
  pdfauthor={Markus Dierigl, Du\v{s}an Novi\v{c}i\'c},
  pdfsubject={axion, 1-form symmetry, gauge group, topological mixing},
  bookmarksopen, bookmarksnumbered, bookmarksopenlevel=2, colorlinks=false, linkcolor=blue, citecolor=blue, urlcolor=blue]{hyperref}

\usepackage{comment}
\usepackage{rotating}
\usepackage{setspace}
\usepackage{longtable}
\usepackage{pdflscape}
\usepackage{caption}

\usepackage{tikz}
\usepackage{tikz-3dplot}
\usetikzlibrary{calc}
\usetikzlibrary{decorations} 
\usepackage{tikz-cd} 

\usepackage[all]{xy}

\numberwithin{equation}{section}

\usepackage[nosort]{cite}

\begin{document}

\baselineskip=18pt

\vspace*{-2cm}
\begin{flushright}
  \texttt{LMU-ASC 14/24}
\end{flushright}

\vspace*{0.6cm} 
\begin{center}
{\Large{\textbf{The Axion is Going Dark}}} \\
 \vspace*{1.5cm}
Markus Dierigl$^1$, Du\v{s}an Novi\v{c}i\'c$^{2}$\\

{
 \vspace*{1.0cm} 
{\it 
${}^1$ Arnold-Sommerfeld-Center for Theoretical Physics,\\
Ludwig-Maximilians-Universit\"at, 80333 M\"unchen, Germany\\ 
${}^2$ Max-Planck-Institut f\"ur Physik, Werner-Heisenberg-Institut, \\ 85748 Garching bei M\"unchen, Germany
}}

\vspace*{0.8cm}
\end{center}
\vspace*{.5cm}

\noindent In this work we explore the effect of a non-Abelian dark sector gauge group that couples to the axion field. In particular, we analyze effects that arise if the dark sector gauge group mixes topologically with the Standard Model. This is achieved by gauging a subgroup of the global center 1-form symmetries which embeds non-trivially in both the visible as well as the dark sector gauge groups. Leaving the local dynamics unchanged, this effect modifies the quantization conditions for the topological couplings of the axion, which enter the estimate of a lower bound on the axion-photon coupling. In the presence of a dark sector this lower bound can be reduced significantly, which might open up interesting new parameter regions for axion physics. We further determine the allowed exotic matter representations in the presence of topological mixing with a dark sector, explore the generalized categorical symmetries of such axion theories, and comment on other model-independent phenomenological consequences.

\thispagestyle{empty}
\clearpage

\setcounter{page}{1}

\newpage
\tableofcontents

\newpage

\section{Introduction}

Axions are one of the most promising candidates for particles beyond the Standard Model. They can provide a dynamical solution for the strong CP problem \cite{Weinberg:1977ma, Peccei:1977hh, Wilczek:1977pj, Peccei:1977ur} and generically contribute to the dark matter density in the universe \cite{Preskill:1982cy, Dine:1982ah, Abbott:1982af}, see also \cite{Adams:2022pbo}. Axions further have a very rich structure of symmetries, including various generalized symmetries, see \cite{Cordova:2022ruw, Brennan:2023mmt, Bhardwaj:2023kri, Schafer-Nameki:2023jdn, Shao:2023gho} for reviews and in particular \cite{Reece:2023czb}. These include the mixture of various higher-form symmetries into higher groups \cite{Hidaka:2020iaz, Brennan:2020ehu, Hidaka:2020izy} as well as non-invertible symmetries \cite{Choi:2022jqy, Cordova:2022ieu, Choi:2022fgx, Yokokura:2022alv}, making axions the ideal laboratory for studying these structures and their physical implications.\footnote{See also \cite{Cordova:2024ypu} for the treatment of a non-invertible Peccei-Quinn symmetry.}

Since axions couple to the gauge dynamics only topologically via the instanton density they are sensitive to subtle differences in the realization of the gauge group. Because of this, they are able to go beyond the implications of the gauge algebra which dictates the local interactions. This is important even for the Standard Model for which the gauge algebra is known to be composed of strong, weak, and hypercharge interactions:
\begin{equation}
\mathfrak{g}_{\text{SM}} = \mathfrak{su}(3) \oplus \mathfrak{su}(2) \oplus \mathfrak{u}(1) \,,   
\end{equation}
but the actual gauge group is not fixed. The allowed realizations of the Standard Model gauge group can be parameterized by 
\begin{equation}
G_{\text{SM}} = \frac{\text{SU}(3) \times \text{SU}(2) \times \text{U}(1)}{\mathbb{Z}_k} \,, \quad \text{with } k \in \{1\,, 2\,, 3\,, 6 \} \,,
\label{eq:SMgauge}
\end{equation}
see, e.g., \cite{Gaiotto:2014kfa, Tong:2017oea}. The value of $k$ leads to quantization conditions for the axion coupling that affects its coupling to the electro-magnetic field after electro-weak symmetry breaking \cite{Choi:2023pdp, Reece:2023iqn, Agrawal:2023sbp, Cordova:2023her}. This quantization leads to a lower bound on the axion-photon coupling which enters various experimental searches and is therefore of great importance to define the experimentally interesting parameter space, see, e.g., \cite{Adams:2022pbo}.

In this work we will demonstrate that the presence of a non-Abelian dark sector can drastically modify the conclusions about a lower bound on the axion-photon coupling. This happens in cases where it mixes topologically with the Standard Model fields, i.e., the total gauge group is given by\footnote{$\widetilde{G}_{\text{d}}$ denotes the simply connected group associated to the algebra $\mathfrak{g}_{\text{d}}$, e.g., SU$(N)$ for $\mathfrak{su}(N)$.}:
\begin{equation}
G = \frac{\text{SU}(3) \times \text{SU}(2) \times \text{U}(1) \times \widetilde{G}_{\text{d}}}{\mathbb{Z}_k} \,.
\label{eq:gauge}
\end{equation}
For non-trivial realizations of $G$ the effective coupling of the axion to the electro-magnetic field $g_{a \gamma \gamma}$ will generically be reduced compared to its Standard Model value, 
\begin{equation}
g^{G}_{a \gamma \gamma} \leq g^{G_{\text{SM}}}_{a \gamma \gamma} \,.
\end{equation}
In certain cases this reduction is significant and might open up new parameter regimes. For example, we find that in the case of a dark sector with $\mathfrak{su}(3)$ algebra and $\mathbb{Z}_6$ quotient the minimal value of the axion photon coupling differs by an order of magnitude in the presence of topological mixing. Specifically,
\begin{equation}
g_{a \gamma \gamma}^{\text{min,SM}} \simeq \frac{\alpha}{2 \pi f} 0.72(4) \gg g_{a \gamma \gamma}^{\text{min,d}} \simeq \frac{\alpha}{2 \pi f} \big( 0.07(6) + \rho \big) \,, 
\end{equation}
where we expect $\rho$, modifying the kinetic mixing of axions with pions in the dark sector, to be sub-leading.

The topological mixing in the presence of a dark gauge sector further has implications on the allowed representations of exotic matter fields, which we determine for general dark sector gauge group. We further consider scenarios in which the dark sector is confining or broken, which leads to different restrictions. This influences other phenomenological applications of the axion field as well and we comment on the corresponding possible effects. These effects can also be constrained by the categorical symmetries of the system which we explore systematically in the presence of a dark sector including anomaly inflow onto the defects of the axion field given by axion domain walls and axion strings. This has the potential to open up interesting possibilities and new corners of the parameter space.

Restricting to a single non-Abelian dark sector gauge group and a single axion field we will quantify the consistency conditions and restrictions depending on the choices of the dark sector gauge algebra $\mathfrak{g}_{\text{d}}$ and the quotient $\mathbb{Z}_k$ with a focus on universal, model-independent properties. For that we use that the quotient by $\mathbb{Z}_k$ in \eqref{eq:gauge} can be understood as gauging part of the global center 1-form symmetries of the theory \cite{Aharony:2013hda, Gaiotto:2014kfa}, which allows for more general gauge backgrounds carrying fractional instanton number \cite{tHooft:1979rtg, vanBaal:1982ag}, see also \cite{Anber:2021upc}. The periodicity of the axion field leads to consistency conditions for these fractional instanton backgrounds which can be phrased in terms of a quantization condition on the topological coupling constants \cite{Cordova:2019jnf, Cordova:2019uob, Choi:2023pdp, Reece:2023iqn, Agrawal:2023sbp, Cordova:2023her}. Fractional instanton configurations in the dark sector gauge group need to be included in the evaluation of these consistency conditions which modifies the result compared to those obtained from the Standard Model alone.

The rest of the manuscript is organized as follows. In Section \ref{sec:mix} we explain the notion of topological mixing, its effect on allowed matter representation, as well as its formulation in terms of the gauging of a subgroup of the center 1-form symmetries. We further explain how this gauging allows for gauge configurations with fractional instanton numbers. In Section \ref{sec:axtop} we determine how the realization of the total gauge group leads to quantization of the topological axion coupling and explore the influence of the presence of a dark sector gauge group on the minimal value of the axion photon coupling. The generalized categorical symmetries and defects of the axion system are discussed in Section \ref{sec:sym} with a focus on the inclusion of a dark sector. In Section \ref{sec:phenocons} phenomenological consequences of coupling axions to the dark sector are discussed. Finally, we point out possible generalizations and conclude in Section \ref{sec:concl}. In Appendix \ref{app:fracinstsub} we discuss the fractional instanton number in cases in which one only mods out a subgroup of the center symmetry.

\section{Topological mixing}
\label{sec:mix}

In this section we discuss the consequences of the topological mixing of gauge groups. For this we note that being given the gauge algebra of a theory does not specify the gauge group, which requires more detailed information, for example the allowed representations or the spectrum of line operators \cite{Aharony:2013hda, Gaiotto:2014kfa, Tong:2017oea}. For a gauge algebra $\mathfrak{g}$ we denote by $\widetilde{G}$ the associated maximal form of the gauge group. For a non-Abelian Lie algebra this is given by the simply-connected realization whereas for U$(1)$ this fixes the charge quantization, see Table \eqref{eq:groups}. The group $\widetilde{G}$ has the maximal center subgroup $\mathcal{Z}_{\widetilde{G}}$, i.e., the subgroup commuting with all group elements. 

As an example consider the special unitary group SU$(3)$, whose center elements form the discrete subgroup $\mathbb{Z}_3$ and are represented by elements of the form
\begin{equation}
\begin{pmatrix} e^{2 \pi i / 3} & 0 & 0 \\ 0 & e^{2 \pi i/3 } & 0 \\ 0 & 0 & e^{2 \pi i/3} \end{pmatrix} = e^{2 \pi i /3} \, \mathbf{1} \in \text{SU}(3) \,.
\end{equation}
Being diagonal matrices, these elements clearly commute with all the SU$(3)$ group elements.

For the groups relevant to us in later sections the Lie algebra $\mathfrak{g}$, the associated maximal Lie group $\widetilde{G}$, and its center subgroup are given by\footnote{Our convention for labeling the symplectic groups is such that $\mathfrak{sp}(1) \simeq \mathfrak{su}(2)$.}
\begin{equation}
\renewcommand*{\arraystretch}{1.5}
\begin{array}{| c | c | c |}
\hline
\text{Lie algebra } \mathfrak{g}  & \text{Lie group } \widetilde{G} & \text{center } \mathcal{Z}_{\widetilde{G}}\\ \hline \hline 
\mathfrak{su}(n) & \text{SU}(n) & \mathbb{Z}_n \\ \hline
\mathfrak{so}(2n+1) & \text{Spin}(2n+1) & \mathbb{Z}_2 \\ \hline
\mathfrak{so}(4n) & \text{Spin} (4n) & \mathbb{Z}_2 \times \mathbb{Z}_2 \\ \hline 
\mathfrak{so}(4n+2) & \text{Spin} (4n + 2) & \mathbb{Z}_4 \\ \hline
\mathfrak{sp}(n) & \text{Sp}(n) & \mathbb{Z}_2 \\ \hline
\mathfrak{e}_6 & \text{E}_6 & \mathbb{Z}_3 \\ \hline
\mathfrak{e}_7 & \text{E}_7 & \mathbb{Z}_2 \\ \hline
\mathbb{R} & \text{U}(1) & \text{U}(1) \\ \hline
\end{array}
\label{eq:groups}
\renewcommand*{\arraystretch}{1}
\end{equation}
The remaining Lie groups E$_8$, G$_2$, and F$_4$ all have trivial center and will not be relevant for our discussion.

In general, a gauge group associated to a single Lie algebra summand $\mathfrak{g}$ takes the form
\begin{equation}
G = \frac{\widetilde{G}}{\mathcal{Z}} \,, \quad \text{with } \mathcal{Z} \subset \mathcal{Z}_{\widetilde{G}} \,,
\end{equation}
where we divide by $\mathcal{Z}$, a subgroup of the center. Whenever the gauge algebra contains more than one summand, the quotient by $\mathcal{Z}$ can affect each gauge group factor simultaneously. We will refer to this phenomenon as topological mixing, since the local dynamics of a theory with such a gauge group remains unchanged. For several gauge group factors $\widetilde{G}_i$ this can be written as
\begin{equation}
G = \frac{\prod_{i} \widetilde{G}_i}{\mathcal{Z}} \,.
\label{eq:globalgroup}
\end{equation}
To specify the resulting gauge group $G$ one further needs to fix the group homomorphisms from $\mathcal{Z}$ to each $\mathcal{Z}_{\widetilde{G}_i}$. These can be inferred from the image of a set of generators of $\mathcal{Z}$, which also includes the possibility of $\mathcal{Z}$ having more than one factor. This data can be captured by an integer $\ell_i$, which specifies the maps
\begin{equation}
\mathcal{Z} \supset \mathbb{Z}_n \rightarrow \mathbb{Z}_{m_i} \subset \mathcal{Z}_{\widetilde{G}_i} \,,
\end{equation}
by the image of the generator
\begin{equation}
1 \text{ mod } n \mapsto \ell_i \text{ mod } m_i \,.
\end{equation}
The subgroup in the image of $\mathbb{Z}_n$ is given by $\mathbb{Z}_{\frac{m_i}{\text{gcd}(\ell_i, m_i)}}$, where $\text{gcd}$ denotes the greatest common divisor. Moreover, for this to be well-defined the generated subgroup cannot be bigger than $\mathbb{Z}_n$, which restricts the allowed values of $\ell_i$ and imposes that $n$ is mapped to a multiple of $m_i$. In particular, it requires that
\begin{equation}
\frac{\ell_i \, n}{m_i} \in \mathbb{Z} \,,
\label{eq:consisthomomorphism}
\end{equation}
where we regard the $\ell_i$ parameter mod $m_i$, such that the identity element in $\mathbb{Z}_n$ is mapped to the identity element in $\mathbb{Z}_m$. Note, that since for $\mathfrak{g} = \mathfrak{so}(4n)$ the center is $\mathbb{Z}_2 \times \mathbb{Z}_2$ one needs to specify maps into both $\mathbb{Z}_2$ factors. If $\widetilde{G}_i = \text{U}(1)$ we can specify the group homomorphism 
\begin{equation}
\mathcal{Z} \supset \mathbb{Z}_n \rightarrow \text{U}(1) = \mathbb{Z}_{\widetilde{G}_i} \,,
\end{equation}
by the map
\begin{equation}
1 \text{ mod } n \mapsto e^{2 \pi i \ell_i / n} \in \text{U}(1) \,,
\label{eq:U1embed}
\end{equation}
which produces a $\mathbb{Z}_{\frac{n}{\text{gcd}(\ell_i,n)}}$ subgroup of $\text{U}(1) = \mathcal{Z}_{\widetilde{G}_i}$.

Let us illustrate this in more detail for four different examples:

\begin{itemize}
    \item{$\mathcal{Z} = \mathbb{Z}_6$ and $\widetilde{G}_i = \text{SU}(3)$: The homomorphisms are specified by the maps of the generator of $\mathbb{Z}_6$ into $\mathcal{Z}_{\widetilde{G}_i} = \mathbb{Z}_3$
    \begin{equation}
    1 \text{ mod } 6 \mapsto \ell_i \text{ mod } 3 \,.
    \end{equation}
     In both non-trivial cases the generated subgroup of $\mathcal{Z}_{\widetilde{G}_i}$ is the full $\mathbb{Z}_3$, but the specific homomorphism differs, captured by $\ell_i \in \{ 1, 2\}$.}
    \item{$\mathcal{Z} = \mathbb{Z}_6$ and $\widetilde{G}_i = \text{SU(12)}$: With the group homomorphism into $\mathcal{Z}_{\widetilde{G}_i} = \mathbb{Z}_{12}$ specified by
    \begin{equation}
    1 \text{ mod } 6 \mapsto \ell_i \text{ mod } 12 \,.
    \end{equation}
    For this to be a group homomorphism $\ell_i$ needs to be even. And we find the generated subgroup and remaining parameters of the non-trivial homomorphism 
    \begin{equation}
    \renewcommand*{\arraystretch}{1.5}
    \begin{array}{| c | c | c |}
    \hline
    \ell_i & \frac{m_i}{\text{gcd}(\ell_i,m_i)} & \mathbb{Z}_{\frac{m_i}{\text{gcd}(\ell_i,m_i)}} \subset \mathcal{Z}_{\widetilde{G}_i} \\ \hline \hline 
    2 & 6 & \mathbb{Z}_6 \\ \hline
    4 & 3 & \mathbb{Z}_3 \\ \hline
    6 & 2 & \mathbb{Z}_2 \\ \hline
    8 & 3 & \mathbb{Z}_3 \\ \hline
    10 & 6 & \mathbb{Z}_6 \\ \hline
    \end{array}
    \renewcommand*{\arraystretch}{1}
    \end{equation}
    We see that there are many different possibilities parametrized by the generated subgroup in $\mathcal{Z}_{\widetilde{G}_i}$.}
    \item{$\mathcal{Z} = \mathbb{Z}_2$ and $\widetilde{G}_i = \text{Spin}(8)$: For this we need to specify two maps, since $\mathcal{Z}_{\widetilde{G}_i} = \mathbb{Z}_2 \times \mathbb{Z}_2$ and we find
    \begin{equation}
    1 \text{ mod } 2 \mapsto (\ell^1_i \text{ mod } 2 \,, \ell^2_i \text{ mod } 2) \,,
    \end{equation}
    allowing for three different non-trivial embeddings, see also Appendix \ref{app:fracinstsub}.}
    \item{$\mathcal{Z} = \mathbb{Z}_6$ and $\widetilde{G}_i = \text{U}(1)$: The map is specified by
    \begin{equation}
    1 \text{ mod } 6 \mapsto e^{2 \pi i \ell_i / 6} \,,
    \end{equation}
    which as for SU$(12)$ leads to various subgroups of U$(1)$ and group homomorphisms. To be specific, one finds $\mathbb{Z}_6$ for $\ell_i \in \{1 \,,5 \}$, $\mathbb{Z}_3$ for $\ell_i \in \{ 2\,, 4 \}$, and $\mathbb{Z}_2$ for $\ell_i = 3$.}
\end{itemize}
For the quotient in \eqref{eq:globalgroup} to lead to topological mixing we demand that $\mathcal{Z}$ embeds non-trivially in a product of at least two different gauge group factors $\widetilde{G}_i$. This quotient by part of the center symmetry has various consequences:
\begin{itemize}
    \item{Restriction of the allowed representations}
    \item{More general gauge theory backgrounds}
    \item{Fractional instantons}
\end{itemize}
We will discuss these in the following subsections and in particular choose 
\begin{equation}
\prod_i \widetilde{G}_i \supset \widetilde{G}_{\text{SM}} = \text{SU}(3) \times \text{SU} (2) \times \text{U}(1) \,,
\end{equation}
the form of the Standard Model gauge group with the largest center symmetry.

First, let us recall that even without additional gauge factors in \eqref{eq:globalgroup} there can be topological mixing, parametrized by the different forms of the Standard Model gauge group $G_{\text{SM}}$ in \eqref{eq:SMgauge}. With the center of $\widetilde{G}_{\text{SM}}$ given by
\begin{equation}
\mathcal{Z}_{\widetilde{G}_{\text{SM}}} = \mathbb{Z}_3 \times \mathbb{Z}_2 \times \text{U}(1) \,.
\end{equation}
The specific embedding of the possible discrete quotients $\mathbb{Z}_k$ with $k \in \{1 \,, 2 \,, 3 \,, 6 \}$, where $\mathbb{Z}_1$ is the trivial group, is given by
\begin{align}
\begin{split}
\mathbb{Z}_3:& \quad 1 \text{ mod } 3 \mapsto \big(1 \text{ mod } 3 \,, 0 \text{ mod } 2 \,, e^{2 \pi i \frac{2}{3}} \big) \in \mathcal{Z}_{\widetilde{G}_{\text{SM}}} \,, \quad (\ell_3 \,, \ell_2 \,, \ell_1) = (1 \,, 0 \,, 2) \,, \\
\mathbb{Z}_2:& \quad 1 \text{ mod } 2 \mapsto \big(0 \text{ mod } 3 \,, 1 \text{ mod } 2 \,, e^{2 \pi i \frac{1}{2}} \big) \in \mathcal{Z}_{\widetilde{G}_{\text{SM}}} \,, \quad (\ell_3 \,, \ell_2 \,, \ell_1) = (0 \,, 1 \,, 1) \,,
\end{split}
\label{eq:SMZ2Z3}
\end{align}
Here, $\ell_3$, $\ell_2$, and $\ell_1$ specify the maps to $\mathcal{Z}_{\text{SU}(3)}$, $\mathcal{Z}_{\text{SU}(2)}$, and $\text{U(1)}$ of $\mathcal{Z}_{\widetilde{G}_{\text{SM}}}$, respectively. The transformation under $\mathbb{Z}_6$ can be deduced by the fact that $\mathbb{Z}_6 = \mathbb{Z}_3 \times \mathbb{Z}_2$ and is generated by, e.g.,
\begin{equation}
\mathbb{Z}_6: \quad 1 \text{ mod } 6 \mapsto \big(1 \text{ mod } 3 \,, 1 \text{ mod } 2 \,, e^{2 \pi i \frac{1}{6}}\big) \in \mathcal{Z}_{\widetilde{G}_{\text{SM}}} \,, \quad (\ell_3 \,, \ell_2 \,, \ell_1) = (1 \,, 1 \,, 1) \,.
\label{eq:SMZ6}
\end{equation}
In the following we further include a single additional non-Abelian gauge group factor $\widetilde{G}_{\text{d}}$ and interpret it in terms of a dark sector gauge group. Here, we focus on a single non-Abelian dark sector in order to avoid the strong additional constraints for dark U$(1)$ gauge fields imposed by kinetic mixing effects, see, e.g., \cite{Fabbrichesi:2020wbt} and \cite{Agrawal:2022lsp}, and not to clutter notation by multiple additional gauge factors. The total gauge group is then given by
\begin{equation}
G = \frac{\widetilde{G}_{\text{SM}} \times \widetilde{G}_\text{d}}{\mathcal{Z}} \,,
\label{eq:gaugegroup}
\end{equation}
where $\mathcal{Z}$ embeds non-trivially in both factors.

We will see that the presence of topological mixing with the dark sector has several important consequences for exotic matter representations as well as in the presence of an axion field. To allow for a sensible phenomenology we further assume a mass gap for the dark sector gauge fields. In particular we will distinguish between the following two scenarios:

\subsubsection*{Confining dark sector}

In this scenario the non-Abelian dark sector gauge group is unbroken at low energies and a mass gap is generated dynamically via confinement. We leave the associated energy scale $\Lambda_{\text{d}}$ unspecified, since it depends on the details of the model, but discuss several interesting consequences below. In order not to alter the interactions of Standard Model particles they need to be singlets under the the dark sector gauge group. This constrains the allowed quotients $\mathcal{Z}$ and can be realized by demanding that $\mathcal{Z}$ embeds into $\mathcal{Z}_{\widetilde{G}_{\text{SM}}}$ as described in \eqref{eq:SMZ2Z3} or \eqref{eq:SMZ6}. Depending on the scale $\Lambda_{\text{d}}$ there can be important modifications to the axion potential induced by the strong coupling dynamics in the dark sector. 

Note that such models also lead to interesting dark matter candidates in terms of dark baryons and glueballs, see, e.g., \cite{Strassler:2006im, Bai:2013xga, Boddy:2014yra}, and are very common in string theory compactifications \cite{Cvetic:2012kj, Halverson:2016nfq, Halverson:2018olu}. 

\subsubsection*{Broken dark sector}

The second possibility is that the dark sector gauge group is broken completely at low energies, which means that in principle the Standard Model gauge fields can transform as non-trivial representations under $\widetilde{G}_{\text{d}}$. This however, leads to a multiplicity of the Standard Model matter fields according to the dimension of the dark sector representation, which needs to be implemented carefully to allow for a reasonable phenomenology in the visible sector. All interactions coming from the dark sector would be suppressed by the breaking scale $\Gamma_{\text{d}}$, which can be very high and therefore does not change the measured interactions. In this case $\mathcal{Z}$ does not have to embed as a $\mathbb{Z}_k$ as in \eqref{eq:SMZ2Z3} and \eqref{eq:SMZ6} discussed above and can be more general. Moreover, the modifications of the axion potential will be suppressed by the breaking scale.

This scenario is typical in Grand Unified theories where the Standard Model embeds into a larger gauge group at high energies, see, e.g., \cite{Agrawal:2022lsp} for a recent discussion. It is also generically realized in phenomenological applications of the heterotic string \cite{Gross:1984dd} with its large 10d gauge groups.

\vspace{0.5cm}

In the following we do not want to analyze individual models, with a certain realization of the matter fields, which are also subject to other consistency constraints such as anomaly cancellation. Instead we focus on the model independent restrictions imposed by topological mixing and argue how they alter the general properties of the theory. 

Before we go into the investigation of the axion dynamics we will explain what the topological mixing implies for the allowed representations of matter fields as well as consistent gauge theory backgrounds.

\subsection{Representations and exotic matter}
\label{subsec:reps}

The quotient by a subgroup of the center symmetry in \eqref{eq:globalgroup} has consequences for the allowed representation of matter fields in the theory. In particular, only representations that are invariant under $\mathcal{Z}$ in \eqref{eq:globalgroup} survive the quotient. The fact that all matter representations of the Standard Model are invariant under $\mathbb{Z}_k$, that acts as described in \eqref{eq:SMZ2Z3} and \eqref{eq:SMZ6}, allows for the different global realizations $G_{\text{SM}} = \widetilde{G}_{\text{SM}} / \mathbb{Z}_k$ of the Standard Model gauge group.

For example consider $k = 6$, all the fields of the Standard Model, as well as the right-handed neutrinos, are invariant under $\mathbb{Z}_6$, e.g., the left-handed quarks transform as
\begin{equation}
Q_L = ( \mathbf{3} \,, \mathbf{2} )_{1/6}: \quad e^{2 \pi i/3} \times (-1) \times e^{2 \pi i / 6} = 1 \,.
\end{equation}
Here, the subscript denotes the hypercharge normalized to be in $\tfrac{1}{6} \mathbb{Z}$ as is common in the literature.\footnote{Note, however, that the $\mathbb{Z}_6 \subset \text{U}(1)$ still acts on the charge $\tfrac{1}{6}$ state as the phase $e^{2 \pi i/6}$ as is expected for an embedding with $\ell_1 = 1$.} However, for $k = 6$ representations of the form 
\begin{equation}
Q_{\text{exotic}} = (\mathbf{3} \,, \mathbf{1})_{0}: \quad e^{2 \pi i/3} \neq 1 \,,
\label{eq:exotic}
\end{equation}
are not invariant under $\mathbb{Z}_6$ and therefore would not be allowed, even though they are well-defined for the maximal gauge group $\widetilde{G}_{\text{SM}}$. Once the quotient by $\mathcal{Z}$ involves a dark sector gauge group the allowed representations change accordingly as we will discuss next.

Including the dark sector gauge group we specify the particle representations as
\begin{equation}
\mathbf{R} = (\mathbf{R}_3 \,, \mathbf{R}_2 \,, \mathbf{R}_{\text{d}})_q \,,
\end{equation}
where the charges of U$(1)_Y$ once more normalized to be elements in $\tfrac{1}{6} \mathbb{Z}$. Next, we define the center charge $r_i$ of a representation $\mathbf{R}_i$ as given by its phase under center transformations 
\begin{equation}\label{eq:centercharge}
\mathcal{Z}_{\widetilde{G}_i} \supset \mathbb{Z}_{m_i}: \quad \mathbf{R}_i \rightarrow e^{\frac{2 \pi i}{m_i} r_i} \, \mathbf{R}_i \,,
\end{equation}
e.g., $r_3 = 1 \text{ mod }3$ for the fundamental representation $\mathbf{3}$ of SU$(3)$. Thus, one obtains a consistency condition for each $\mathbb{Z}_n \subset \mathcal{Z}$ factor, which can be phrased in terms of the $\ell_i$ as 
\begin{equation}
\mathcal{Z} \supset \mathbb{Z}_n: \quad \mathbf{R} \rightarrow \text{exp}\bigg( 2 \pi i \sum_i \frac{\ell_i}{m_i} r_i + 2 \pi i \frac{\ell_1}{n}6q \bigg) \mathbf{R} = \mathbf{R} \,,
\end{equation}
which can be rephrased as
\begin{equation}
\sum_i \frac{\ell_i}{m_i} r_i + \frac{\ell_1}{n}6q \in \mathbb{Z} \,.
\end{equation}
As usual these formulas must be modified if $\widetilde{G}_i = \text{Spin}(4n)$, since then $\mathcal{Z}_{\widetilde{G}_i} = \mathbb{Z}_2 \times \mathbb{Z}_2$ and one needs to specify two charges $(r_i^1 \,, r_i^2)$, and two embedding parameters $(\ell_i^1 \,, \ell_i^2)$ as discussed above.

Let us exemplify that for the left-handed quarks $Q_L$ that additionally behave as singlets under the dark sector and $\mathcal{Z} = \mathbb{Z}_6$ embedded as in \eqref{eq:SMZ6}, i.e. $(\ell_3 \,, \ell_2 \,, \ell_1 \,, \ell_{\text{d}}) = (1 \,, 1 \,, 1 \,, 0)$. One finds
\begin{equation}
\sum_i \tfrac{\ell_i}{m_i} r_i + \tfrac{\ell_1}{n}6q = \tfrac{1}{3} + \tfrac{1}{2} + \tfrac{1}{6} + 0 = 1 \in \mathbb{Z} \,,
\end{equation}
which indeed satisfies the quantization condition. If instead we consider $Q_{\text{exotic}}$ in \eqref{eq:exotic}, again transforming as singlet under the dark sector gauge group, one finds
\begin{equation}
\sum_i \tfrac{\ell_i}{m_i} r_i + \tfrac{\ell_1}{n}6q = \tfrac{1}{3} + 0 + 0 + 0 = \tfrac{1}{3} \notin \mathbb{Z} \,,
\end{equation}
which does not satisfy the quantization condition.

Next, we introduce topological mixing with the dark sector and see how this changes the allowed matter representations. In case the dark sector confines we demand that all Standard Model fields are singlets under $\widetilde{G}_{\text{d}}$, i.e., 
\begin{equation}
\mathbf{R}_{\text{SM}} = \big(\mathbf{R}_{3} \,, \mathbf{R}_{2} \,, \mathbf{1}_\text{d} \big)_{q} \,,
\end{equation}
since otherwise they might be confined in dark baryons. This is automatically satisfied if we demand that $\mathcal{Z}$ embeds into $\mathcal{Z}_{\widetilde{G}_{\text{SM}}}$ as the $\mathbb{Z}_k$ groups defined in \eqref{eq:SMZ2Z3} and \eqref{eq:SMZ6}, and there is a non-trivial homomorphism of $\mathbb{Z}_k$ to $\mathcal{Z}_{\text{d}}$ to generate the topological mixing. This allows for the appearance of more general representations. 

For example for $\mathcal{Z} = \mathbb{Z}_6$ and starting with $Q_{\text{exotic}}$ in \eqref{eq:exotic} under the Standard Model one has
\begin{equation}
\mathbf{R} = (\mathbf{3} \,, \mathbf{1} \,, \mathbf{R}_{\text{d}})_0 : \quad \sum_i \tfrac{\ell_i}{m_i} r_i + \tfrac{\ell_1}{n}6q = \tfrac{1}{3} + 0 + 0 + \tfrac{\ell_{\text{d}}}{ m_{\text{d}}} r_{\text{d}} \in \mathbb{Z} \,,
\end{equation}
from which one finds the consistency constraint
\begin{equation}
\tfrac{\ell_{\text{d}}}{ m_{\text{d}}} r_{\text{d}} = \tfrac{2}{3} \text{ mod } 1 \,.
\end{equation}
This can be achieved for example for $\mathfrak{g}_{\text{d}} = \mathfrak{su}(3)$ and the embedding specified by
\begin{equation}
(\ell_3 \,, \ell_2 \,, \ell_1 \,, \ell_{\text{d}}) = (1 \,, 1\,, 1 \,, \ell_{\text{d}}) \,,
\label{eq:exampleembedconf}
\end{equation}
for which we find
\begin{equation}
\mathbf{R} = (\mathbf{3} \,, \mathbf{1} \,, \mathbf{3})_0: \quad \quad \sum_i \tfrac{\ell_i}{m_i} r_i + \tfrac{\ell_1}{n}6q = \tfrac{1}{3} + 0 + 0 + \tfrac{\ell_{\text{d}}}{ 3} \times 1 \in \mathbb{Z} \,,
\end{equation}
which is allowed for $\ell_{\text{d}} = 2$. Similiarly,
\begin{equation}
\mathbf{R} = (\mathbf{3} \,, \mathbf{1} \,, \overline{\mathbf{3}})_0: \quad \sum_i \tfrac{\ell_i}{m_i} r_i + \tfrac{\ell_1}{n}6q = \tfrac{1}{3} + 0 + 0 + \tfrac{\ell_{\text{d}}}{ 3} \times 2 \in \mathbb{Z} \,,
\end{equation}
which is allowed for $\ell_{\text{d}} = 1$.

For the broken dark sector scenario we do not have to demand that the Standard Model fields are singlets under $\widetilde{G}_{\text{d}}$ and one can take more general quotients. The allowed matter representations of the UV gauge group will still be influenced by the topological mixing as discussed above.

For that let us consider another example with $\mathcal{Z} = \mathbb{Z}_6$ and $\mathfrak{g}_{\text{d}} = \mathfrak{su}(3)$, but this time the embedding differs from \eqref{eq:SMZ6} in the Standard Model factors and instead is given by
\begin{equation}
(\ell_3 \,, \ell_2 \,, \ell_1 \,, \ell_{\text{d}}) = (1 \,, 1 \,, -1 \,, 1) \sim (1 \,, 1 \,, 5 \,, 1) \,.
\label{eq:exampleembedbrok}
\end{equation}
In this case the the left-handed quarks need to transform non-trivially under $\widetilde{G}_{\text{d}}$ in order to absorb the phase
\begin{equation}
Q_L = (\mathbf{3}, \mathbf{2})_{1/6}: \quad Q_L \rightarrow e^{2 \pi i \frac{2}{3}} \, Q_L \,,
\end{equation}
under $\mathcal{Z} = \mathbb{Z}_6$. This can be ensured by transforming as $\mathbf{3}$ under $\widetilde{G}_{\text{d}} = \text{SU}(3)$, leading to the consistency condition
\begin{equation}
\mathbf{R} = (\mathbf{3} \,, \mathbf{2} \,, \mathbf{3})_{1/6}: \quad \sum_i \tfrac
{\ell_i}{m_i} r_i + \tfrac{\ell_1}{n}{6}q= \tfrac{1}{3} + \tfrac{1}{2} - \tfrac{1}{6} + \tfrac{1}{3} = 1 \in \mathbb{Z} \,,
\end{equation}
being satisfied. Note that the exotic matter representation \eqref{eq:exotic}, realized via $(\mathbf{3} \,, \mathbf{1} \,, \overline{\mathbf{3}})_0$ is also allowed for this particular embedding.

Similar conclusions about the realization of allowed exotic matter states can be applied for all the possible gauge groups and follow the same logic as explained above. This demonstrates that the global form of the gauge group leads to significant restrictions on the allowed exotic matter states, see also the analysis in \cite{Li:2024nuo}.

\subsection{Non-trivial 1-form symmetry backgrounds}
\label{subsec:1gauging}

 One can interpret taking the quotient in \eqref{eq:globalgroup} as gauging part of the center 1-form symmetry \cite{Aharony:2013hda, Gaiotto:2014kfa}. This center 1-form symmetry acts on Wilson lines labeled by a representation $\mathbf{R}$ of the gauge group with the charge $r$ defined as in \eqref{eq:centercharge}
\begin{equation}
\mathcal{W}_{\mathbf{R}} = \text{tr} \Big( P \, \text{exp} \Big( i \oint A_{\mathbf{R}} \Big) \Big) \,,
\end{equation}
where $P$ denotes a path-ordering in the exponential. Since the charged objects are one-dimensional, the associated symmetry is called a 1-form symmetry. The existence of dynamical particles transforming under the center breaks this symmetry explicitly, since now the Wilson lines can end on a charged local operator. For the Standard Model the discrete $\mathbb{Z}_6$ is not broken and one can gauge (subgroups of) it. Once one includes the dark sector gauge group the gauging and breaking depends on the specific matter representations.\footnote{In quantum gravity it is expected that all global symmetries are either gauged or broken, e.g., \cite{Banks:2010zn, Harlow:2018tng}, so in our scenario this would mean that either there are exotic matter representation that break the 1-form symmetry or it necessarily needs to be gauged.}

To perform such a gauging we couple the 1-form center symmetry of $\widetilde{G} = \widetilde{G}_{\text{SM}} \times \widetilde{G}_{\text{d}}$ given by
\begin{equation}
\mathcal{Z}_{\widetilde{G}} = \mathbb{Z}_3 \times \mathbb{Z}_2 \times \text{U}(1) \times \mathcal{Z}_{\text{d}} \,,
\end{equation}
to background gauge fields, which are discrete 2-form fields $B_i$. We will further make the assumption that $\mathcal{Z}$ maximally embeds as $\mathbb{Z}_6$ in the Abelian U$(1)_Y$ of the hypercharge sector. Reason is that one otherwise generically ends up with fractionally charged particles that are additionally charged under the dark sector and are experimentally heavily constrained, see, e.g., \cite{Davidson:2000hf} but also the analysis in \cite{Koren:2024xof}, similar to kinetic mixing with another Abelian factor that can also produce such states \cite{Fabbrichesi:2020wbt}. In this case the background fields are given by elements
\begin{equation}
(B_3 \,, B_2 \,, B_1 \,, B_{\text{d}}) \in H^2 (M; \mathbb{Z}_3) \times H^2 (M; \mathbb{Z}_2) \times H^2 (M; \mathbb{Z}_6) \times H^2 (M; \mathcal{Z}_{\text{d}}) \,,
\end{equation}
where $M$ denotes the spacetime manifold. The gauging proceeds by promoting a linear combination of these background fields to be dynamical, which for a discrete fields merely means that we sum over its configurations in the partition function. The specific form of linear combination which is gauged is determined by the embedding of $\mathcal{Z}$ into the individual factors of $\mathcal{Z}_{\widetilde{G}}$, i.e., by the embedding parameters $\ell_i$.

For $\mathcal{Z} = \mathbb{Z}_6$ with embedding parameters $(\ell_3 \,, \ell_2 \,, \ell_1 \,, \ell_{\text{d}})$ we can parametrize the `dynamical' linear combination by summing over $\omega \in H^2 (M; \mathbb{Z}_6)$ and identifying
\begin{equation}
(B_3 \,, B_2 \,, B_1 \,, B_{\text{d}}) = \Big(\ell_3 \, \omega_3 \,, \ell_2 \, \omega_2 \,, \ell_1 \, \omega \,,  \ell_{\text{d}} \, \omega_{\text{d}} \Big) \,.
\label{eq:background2form}
\end{equation}
Here, $\omega_i$ is the reduction of $\omega$ modulo the order of the different factors of $\mathcal{Z}_{i}$. For embedding of subgroups into U$(1)$ and $\mathcal{Z}_i$ we use the fact that we can rephrase a background in $H^2(M; \mathbb{Z}_k)$ as a background in $H^2(M; \mathbb{Z}_{m k})$ by multiplication by $m$, see Appendix \ref{app:fracinstsub}. Again, we note the usual caveats in case $\mathfrak{g}_{\text{d}} = \mathfrak{so}(4n)$, for which one needs to specify two independent background gauge fields in $H^2(M;\mathbb{Z}_2)$.  

For example, for \eqref{eq:exampleembedconf} one finds the linear combination
\begin{equation}
(B_3 \,, B_2 \,, B_1 \,, B_{\text{d}}) = (\omega_3 \,, \omega_2 \,, \omega \,, \ell_{\text{d}} \, \omega_3) \,.
\label{eq:gaugebackg}
\end{equation}
Similarly, for \eqref{eq:exampleembedbrok} the appropriate linear combination is given by
\begin{equation}
(B_3 \,, B_2 \,, B_1 \,, B_{\text{d}}) = (\omega_3 \,, \omega_2 \,, - \omega \,, \omega_3) \,,
\end{equation}
where one could equally well use $5 \omega$ instead of $- \omega$, since $6 \omega$ is trivial in $H^2(M;\mathbb{Z}_6)$.

Since for the maximal gauge group $\widetilde{G}$ we effectively need to set all $\omega_i$ to zero we see that the gauge group $G = \widetilde{G}/\mathcal{Z}$ has more gauge configurations than $\widetilde{G}$.\footnote{This can also be seen from classifying the gauge configurations via transition functions where in $\widetilde{G}$ the co-cycle condition on triple overlaps needs to produce the identity in $\widetilde{G}$, it can be a non-trivial element in $\mathcal{Z}$ for $\widetilde{G}/\mathcal{Z}$, see e.g. \cite{Gaiotto:2014kfa, Kapustin:2014gua}.} These new gauge field backgrounds, however, lead to a restriction on allowed matter representations, which have to be compatible with the gauging procedure and precisely lead to the restrictions discussed in Section \ref{subsec:reps}. Moreover, after gauging part of the center 1-form symmetry the quantization of instanton numbers is modified as we will discuss next.

\subsection{Instanton numbers}
\label{subsec:instnumb}

Instantons are topological gauge field configurations with a non-trivial value of the integral 
\begin{equation}
n_i = \frac{1}{8 \pi^2} \int \text{tr}(F_i \wedge F_i) \,,
\label{eq:nAinstanton}
\end{equation}
where $F$ is the non-Abelian field strength of $\mathfrak{g_i}$ in differential form notation, see, e.g., \cite{Reece:2023czb}. Since they are characteristic classes of the gauge theory their number, $n_i$, is quantized.\footnote{They are the pullbacks of non-trivial elements in group cohomology with integer coefficients, i.e., originate from $H^4 (BG; \mathbb{Z})$, with classifying space $BG$.} In our conventions of the trace the instanton numbers for simply-connected non-Abelian Lie groups on 
Spin manifolds are integers. For Abelian gauge groups the instanton number is given by
\begin{equation}
n_1 = \frac{1}{8 \pi^2} \int F_1 \wedge F_1 \,,
\label{eq:Abelinst}
\end{equation}
with hypercharge U$(1)_Y$ field strength $F_1 = d A_1$. Once more $n_1$ is an integer on Spin manifolds. It is important to stress that the spacetime is Spin, since four-dimensional Spin manifolds have an even intersection form for $H_2(M;\mathbb{Z})$. This, together with the charge quantization condition $[F_1/2\pi] \in H^2(M;\mathbb{Z})$, is why we can divide by an additional factor of $\tfrac{1}{2}$ in \eqref{eq:Abelinst}. For non-Spin manifolds, e.g., $M = \mathbb{CP}^2$, this is generally not the case and \eqref{eq:Abelinst} can be half-integer. However, this is modified in the presence of non-trivial background fields $B_i$ for the center 1-form symmetries.

In the case of a U$(1)$ gauge symmetry and $\mathcal{Z} = \mathbb{Z}_n$ this can be understood as follows. The background field $B_1 \in H^2(M; \mathbb{Z}_n)$ can be rephrased as a flat 2-form gauge field $b_1 \in H^2(M; \text{U}(1))$ which is roughly related to $B_1$ as
\begin{equation}
b_1 = \tfrac{2 \pi}{n} B_1 \,.
\label{eq:disccont}
\end{equation}
In particular this means that if $B_1$ integrates to $q \in \mathbb{Z}$ mod $n$ on a certain 2-cycle, $b_1$ integrates to $2 \pi q/n$ mod $2\pi$. In the presence of this background field the instanton number is shifted to
\begin{equation}
n_1 = \frac{1}{8 \pi^2} \int (F_1 - b_1) \wedge (F_1 - b_1) \,.
\label{eq:Abinstanton}
\end{equation}
In general this gives rise to fractional instanton contributions of the form 
\begin{equation}
\frac{1}{4 \pi^2} \int F_1 \wedge b_1 \in \tfrac{1}{n} \mathbb{Z} \,, \quad \text{and} \quad \frac{1}{8 \pi^2} \int b_1 \wedge b_1 \in \tfrac{1}{n^2} \mathbb{Z} \,.
\label{eq:U1frac}
\end{equation}
At the same time these new configurations violate the quantization condition for the magnetic flux $F_1$ and would require electric charges to be multiples of $n$, which is precisely the U$(1)$ version of the restriction of representations discussed in Section \ref{subsec:reps}. Equivalently, since the action is quadratic in $F_1$ these results can been seen explicitly by noticing that gauging the $\mathbb{Z}_n$ 1-form symmetry effectively replaces $F_1$ with $\tfrac{1}{n} F_1$ in the Lagrangian. 

A similar conclusion holds for non-Abelian gauge groups where the non-trivial 1-form symmetry backgrounds $B_i$ lead to fractional contributions to the instanton numbers. To extract their fractionality one modifies the field strength to include contributions of $B_i$ and evaluates their effect. This has been done in detail in \cite{Aharony:2013hda, Kapustin:2014gua, Gaiotto:2017yup, Cordova:2019uob} to which we refer for a detailed derivation. 

Here, we briefly recall the derivation in the special case $\widetilde{G} = \text{SU}(n)$ and $\mathcal{Z} = \mathbb{Z}_k$ and embedding specified by
\begin{equation}
1 \text{ mod } k \mapsto \ell \text{ mod } n \,.
\end{equation}
For the resulting gauge group to be SU$(n)/\mathbb{Z}_k$ we further have to demand the stronger condition compared to \eqref{eq:consisthomomorphism} that $k$ divides $n$ to generate a proper subgroup, where in general we would allow for a subgroup of $\mathcal{Z}$ to map non-injectively to $\mathcal{Z}_{\widetilde{G}_i}$. The non-trivial background field $B \in H^2(M; \mathbb{Z}_n)$ is defined by
\begin{equation}
B = \ell \, \omega \, \mod n,
\end{equation}
where $\omega \in H^2(M;\mathbb{Z}_k)$, is the parameter we sum over in the partition function, and $B$ is well defined because of \eqref{eq:consisthomomorphism}. One also has
\begin{equation}
\tfrac{n}{\text{gcd}(\ell,n)} = k \,.
\end{equation}
Because of this there is a way to interpret $B$ as a background in $H^2(M;\mathbb{Z}_k)$ as one would expect from a quotient by $\mathbb{Z}_k$, see Appendix \ref{app:fracinstsub}. To see how this non-trivial background $B$ modifies the instanton number we promote the SU$(n)$ field strength $F$ to a U$(n)$ field strength $F'$, see \cite{Kapustin:2014gua, Gaiotto:2017yup}, which can be done by introducing a U$(1)$ gauge field $f = da$ and defining
\begin{equation}\label{eq:ungauge}
F' = F + \tfrac{1}{n} f \, \mathbf{1} \,,
\end{equation}
with unit matrix $\mathbf{1}$. One then identifies the U$(1)$ part with the background gauge field, which after using the identification \eqref{eq:disccont} in terms of $b \in H^2(M;\text{U}(1))$ takes the form
\begin{equation}
\text{tr}(F') = f = n  \, b \,.
\end{equation}
This can be implemented via a Lagrange multiplier in the action, see \cite{Kapustin:2014gua, Gaiotto:2017yup}. One then replaces the instanton density \eqref{eq:nAinstanton} with
\begin{equation}
\frac{1}{8 \pi^2} \int \text{tr} \big( (F' -  \, b \, \mathbf{1}) \wedge (F' -  \, b \, \mathbf{1}) \big) \,,
\label{eq:nAinstantonmod}
\end{equation}
which contains the U$(n)$ field and the background $b$. Since the second Chern class of $F'$ integrates to integer values
\begin{equation}
\int c_2(F') = \int \bigg( \frac{1}{8 \pi^2} \text{tr}(F' \wedge F') - \frac{1}{8 \pi^2} \text{tr}(F') \wedge \text{tr}(F') \bigg) \in \mathbb{Z} \,,
\end{equation}
we identify the fractional part of \eqref{eq:nAinstantonmod} as the second term in 
\begin{equation}
\frac{1}{8 \pi^2} \int \text{tr} \big( (F' -   b \, \mathbf{1}) \wedge (F' -   b \, \mathbf{1}) \big) = \int c_2 (F') + \frac{1}{8 \pi^2} \int  n(n-1) \, b \wedge b \,.
\end{equation}
Going back to the discrete gauge field $\omega$, which is summed over,  using $b=\frac{2\pi}{n} B=\frac{2\pi}{n} \ell \omega$ this can be written as
\begin{equation}
n^{\text{frac}} =  \frac{(n-1)}{2 n} \int \mathcal{P}(\ell \omega) \,.
\label{eq:fracexamp}
\end{equation}
Here, $\mathcal{P}$ is the Pontryagin square, see, e.g., \cite{Kapustin:2013qsa}, which is a cohomological operation
\begin{equation}
\mathcal{P}: \quad H^2 (M; \mathbb{Z}_n) \rightarrow H^4 (M; \mathbb{Z}_{p}) \,,
\end{equation}
where $p=n$ for odd $n$ and $p=2n$ for even $n$. It refines the bilinear form given by the cup product
\begin{equation}
\cup: H^2(M;\mathbb{Z}_n) \times H^2(M;\mathbb{Z}_n) \rightarrow H^{4} (M;\mathbb{Z}_n) \,.
\end{equation}
of $B$ with itself. It is only relevant for $n$ even, since for $n$ odd one has $\mathcal{P}(B) = B \cup B \in H^4(M;\mathbb{Z}_n)$. For even $n$ it introduces an additional factor of $\tfrac{1}{2}$ in the fractional part of the instantons. In cases where $B \in H^2(M;\mathbb{Z}_n)$ has a well-defined lift to integer cohomology $B_{\mathbb{Z}} \in H^2(M;\mathbb{Z})$, which we will always assume and which is the case if $H^3(M;\mathbb{Z})$ is torsion free\footnote{In case this lift does not exist, one can still perform the calculations above using the definition of the Pontryagin square on the co-chain level, see \cite{Kapustin:2013qsa}. \label{foot:cochain}}, it is given by $\mathcal{P}(B) = B_{\mathbb{Z}} \cup B_{\mathbb{Z}}$ mod $2n$.

For example for $G = \text{SU}(2)/\mathbb{Z}_2$, where $\ell = 1$ since we assume a non-trivial embedding, the fractional instanton number in the presence of $B \in H^2(M;\mathbb{Z}_2)$ is given by
\begin{equation}
n_2^{\text{frac}} = \tfrac{1}{4} \int \mathcal{P} (B) = \tfrac{1}{4} \int B_{\mathbb{Z}} \cup B_{\mathbb{Z}} \text{ mod } 1 \enspace \in  \{ 0 \,, \tfrac{1}{2} \} \,,
\end{equation}
where we once more used the evenness of the intersection form on Spin manifolds.

For other simple Lie algebras in the presence of the background $B \in H^2(M; \mathcal{Z}_{\widetilde{G}})$ one can determine fractional instanton contributions by utilizing different embeddings of SU$(n)$ groups \cite{Aharony:2013hda, Cordova:2019uob}. The result is presented in the table:
\begin{equation}
\renewcommand*{\arraystretch}{1.5}
\begin{array}{| c | c |}
\hline
\text{gauge algebra } \mathfrak{g} & \text{fractional instanton contribution for } G = \widetilde{G}/\mathcal{Z}_{\widetilde{G}} \\ \hline \hline
\mathfrak{su}(n) & \tfrac{n-1}{2n} \int \mathcal{P}(B) \\ \hline
\mathfrak{so}(2n+1) & \tfrac{1}{2} \int \mathcal{P}(B) \\ \hline
\mathfrak{so}(4n) & \tfrac{n}{4} \int \mathcal{P}(B_L + B_R) + \tfrac{1}{2} \int B_L \cup B_R \\ \hline
\mathfrak{so}(4n+2) & \frac{2n+1}{8} \int \mathcal{P}(B) \\ \hline
\mathfrak{sp}(n) & \tfrac{n}{4} \int \mathcal{P}(B) \\ \hline
\mathfrak{e}_6 & \tfrac{2}{3} \int \mathcal{P}(B) \\ \hline
\mathfrak{e}_7 & \tfrac{3}{4} \int \mathcal{P}(B) \\ \hline
\end{array}
\label{eq:fracinst}
\renewcommand*{\arraystretch}{1}
\end{equation}
Indeed, setting $\ell =1$, and therefore gcd$(\ell,n) = 1$ and $n = k$, in \eqref{eq:fracexamp}, we find
\begin{equation}
n^{\text{frac}} = \frac{n-1}{2n} \int \mathcal{P}(B) \,,
\end{equation}
which conincides with the table. 

In case $\mathcal{Z}$ is only a subgroup of the center of the dark sector $\widetilde{G}_{\text{d}}$, i.e., if $\text{gcd}(\ell_{\text{d}}, m_{\text{d}}) \neq 1$, this affects that fractionality condition. Note that this is only relevant for SU$(n)$ with $n$ not prime and Spin$(2n)$, since otherwise the center does not have non-trivial subgroups. We discuss the various cases in Appendix \ref{app:fracinstsub}.

In the models above the overall quotient $\mathcal{Z}$ embeds into several gauge factors. As we discussed in Section \ref{subsec:1gauging} this correlates the various background fields for the individual 1-form center symmetries. It also implies that fractional instanton contributions in one gauge factor necessarily are accompanied by fractional contributions in a different sector. For $\mathcal{Z} = \mathbb{Z}_6$ and $\mathfrak{g}_{\text{d}} = \mathfrak{su}(3)$ with embedding \eqref{eq:exampleembedconf} the fractional parts are given by
\begin{align}
\begin{split}
n_3^{\text{frac}} &= \tfrac{1}{3} \int \mathcal{P}(\omega_3) \,, \enspace n_2^{\text{frac}} = \tfrac{1}{4} \int \mathcal{P}(\omega_2) \,, \enspace n_{\text{d}}^{\text{frac}} = \tfrac{\ell_{\text{d}}^2}{3} \int \mathcal{P}(\omega_3) \,, \\
n_1^{\text{frac}} &= - \tfrac{1}{6} \int c_1 (F_1) \cup \omega + \tfrac{1}{72} \int \mathcal{P}(\omega) \,,  
\end{split}
\end{align}
where $c_1(F_1) \in H^2(M;\mathbb{Z})$ is the first Chern class of the hypercharge gauge field represented by $[\tfrac{1}{2\pi} F_1]$. The different embedding \eqref{eq:exampleembedbrok} leads to positive sign for the first term for $n_1^{\text{frac}}$, but is otherwise identical after setting $\ell_{\text{d}} = 1$.

In the next section we will see how the fractional instanton charges induced by the quotient $\mathcal{Z}$ enters the dynamics of axion physics.

\section{Axions and topological mixing}
\label{sec:axtop}

Now that we have analyzed how a topological mixing influences the possible representations and leads to more general gauge backgrounds that can carry fractional instanton numbers we analyze how this influences the behavior of an axion field.

For us, an axion field $a$ is a real, pseudoscalar particle that is periodic
\begin{equation}
a \sim a + 2 \pi \,,
\label{eq:axperiod}
\end{equation}
Its kinetic term in differential form notation, see \cite{Reece:2023czb} for an introduction, is given by
\begin{equation}\label{eq:axkin}
\mathcal{L}^a_{\text{kin}} = \tfrac{1}{2} f^2 da \wedge \ast da \,,
\end{equation}
with dimension-full parameter $f$ known as the axion decay constant. The canonically quantized axion field, of mass dimension 1, would therefore be $fa$ and have periodicity $2 \pi f$. Importantly, the axion couples topologically to the instanton density of the gauge groups, for which we include the dark sector. This coupling is given by
\begin{equation}\label{eq:axcoupling}
\mathcal{L}^a_{\text{top}} = \frac{i a}{8 \pi^2} \big( \kappa_3 \, \text{tr}(F_3 \wedge F_3) + \kappa_2 \, \text{tr} (F_2 \wedge F_2) + \kappa_1 \, F_1 \wedge F_1 + \kappa_{\text{d}} \, \text{tr}(F_{\text{d}} \wedge F_{\text{d}}) \big) \,,
\end{equation}
with the topological coupling constants $\kappa_i$. This interaction must not violate the periodicity condition for the axion \eqref{eq:axperiod}, which can be interpreted as a gauged shift symmetry. More specifically, thinking of $a$ as a real field it has $\mathbb{R}$ shift symmetry and by gauging a discrete subgroup $2\pi \mathbb{Z}\subset \mathbb{R}$ one imposes a periodicity condition. 

For $G = \widetilde{G}_{\text{SM}} \times \widetilde{G}_{\text{d}}$, i.e., trivial quotient, this simply implies that the constants $\kappa_i$ are all integers since all instantons are integer quantized. However, once there is non-trivial topological mixing of the gauge theory factors, there are fractional contributions to the instanton numbers, leading to a correlation of the quantization condition of the coupling constants $\kappa_i$. In \cite{Choi:2023pdp, Reece:2023iqn, Agrawal:2023sbp} this was used to show that the axion-photon coupling takes a minimal value that depends on the topological mixing in the Standard Model. Here, we include the influence of mixing with a dark sector gauge group and find that the constraints are modified.

\subsection{Quantization of topological terms}
\label{subsec:quant}

In the following we analyze the quantization condition for the constants $\kappa_i$ in the presence of non-trivial topological mixing involving a non-Abelian dark sector gauge group. Thus, we assume a gauge group of the form 
\begin{equation}
G = \frac{\widetilde{G}_{\text{SM}} \times \widetilde{G}_{\text{d}}}{\mathcal{Z}} \,,
\end{equation}
with a choice of embedding specified by $(\ell_3 \,, \ell_2 \,, \ell_1 \,, \ell_{\text{d}})$. As discussed in Section \ref{subsec:1gauging} this embedding identifies a linear combination of the 1-form symmetry background fields which are summed over in the partition function, gauging the corresponding subgroup of the center 1-form symmetries. Assuming for simplicity that $\mathcal{Z} = \mathbb{Z}_k$ has a single factor, which can be easily generalized, the dynamical gauge field takes the general form
\begin{equation}
(B_3 \,, B_2 \,, B_1 \,, B_{\text{d}}) = \Big(\ell_3 \, \omega_3 \,, \ell_2 \, \omega_2 \,, \ell_1 \, \omega \,, \ell_d \, \omega_{\text{d}} \Big) \,,
\label{eq:backg}
\end{equation}
with $\omega \in H^2(M;\mathcal{Z})$ and $\omega_i$ its mod $m_i$ reductions as discussed around \eqref{eq:background2form}. The last entry $\omega_{\text{d}}$ is the reduction of $\omega$ mod the order of $\mathcal{Z}_{\text{d}}$, with the usual caveat for Spin$(4n)$ with its two factors of the center. With this one can evaluate the fractional contribution to the instanton number and their coupling to the axion 
\begin{equation}
\mathcal{L}^a_{\text{top,frac}} = a \big( \alpha_3 \, \mathcal{P} (\omega_3) + \alpha_2 \, \mathcal{P}(\omega_2) + 
\alpha_1 \, \mathcal{P}(\omega) - \widetilde{\alpha}_1 \, \omega \cup c_1(F_1) + \alpha_{\text{d}} \, \mathcal{P}(\omega_{\text{d}}) \big)
\label{eq:axfrac}
\end{equation}
The fractional prefactors $\alpha_i$ depend on the specific form of the embedding and are given by
\begin{equation}
\alpha_3 = \tfrac{\ell_3^2}{3} \, \kappa_3 \text{ mod } 1 \,, \quad \alpha_2 = \tfrac{\ell_2^2}{4} \, \kappa_2 \text{ mod } 1 \,, \quad \alpha_1 = \tfrac{\ell_1^2}{2 k^2} \, \kappa_1 \text{ mod } 1 \,, \quad \widetilde{\alpha}_1 = \tfrac{\ell_1}{k} \, \kappa_1 \text{ mod } 1 \,,
\label{eq:fraccoefficients}
\end{equation}
and for the dark sector
\begin{equation}
\alpha_{\text{d}} = c^{\text{frac}} \ell_{\text{d}}^2 \, \kappa_{\text{d}} \,,
\end{equation}
where $c^{\text{frac}}$ is the fractional coefficient in Table \eqref{eq:fracinst}, which depends on the choice of $\mathfrak{g}_{\text{d}}$. Since the axion periodicity is fixed to $2\pi$ the coupling to the topological quantities in \eqref{eq:axfrac} has to respect that.\footnote{If this is not the case this can be interpreted as a mixed anomaly between the gauged $2\pi \mathbb{Z}$ shift symmetry of the axion and the gauged center 1-form symmetries presenting an inconsistency \cite{Cordova:2019jnf, Cordova:2019uob}. This is also common for topological terms involving higher form fields in larger dimensions, see, e.g., \cite{Apruzzi:2020zot, BenettiGenolini:2020doj, Cvetic:2020kuw, Cvetic:2021sxm}.} This leads to the quantization condition
\begin{equation}
\int \big( \alpha_3 \, \mathcal{P} (\omega_3) + \alpha_2 \, \mathcal{P}(\omega_2) + 
\alpha_1 \, \mathcal{P}(\omega) - \widetilde{\alpha}_1 \, \omega \cup c_1 (F_1) + \alpha_{\text{d}} \, \mathcal{P}(\omega_{\text{d}}) \big) \in \mathbb{Z} \,,
\label{eq:axionquant}
\end{equation}
for all backgrounds of the form \eqref{eq:backg} which are summed over. However, note that not the individual pieces but only their linear combination has to satisfy this constraint.

Let us first set $\alpha_{\text{d}}$ to zero and reproduce the known result for $\mathcal{Z} = \mathbb{Z}_6$, $(\ell_3 \,, \ell_2 \,, \ell_1) = (1 \,, 1 \,, 1)$, in the Standard Model. One has
\begin{equation}
\int \big( \tfrac{1}{3} \kappa_3 \, \mathcal{P}(\omega_3) + \tfrac{1}{4} \kappa_2 \, \mathcal{P}(\omega_2) + \tfrac{1}{72} \kappa_1 \, \mathcal{P}(\omega) - \tfrac{1}{6} \kappa_1 \, \omega \cup c_1(F_1) \big) \in \mathbb{Z} \,.
\label{eq:SMfracaxion}
\end{equation}
Note that for this condition to be satisfied $\kappa_1$ needs to be a multiple of $6$ and one has
\begin{equation}
\kappa_1 \in 6 \mathbb{Z} \,,
\end{equation}
which makes the integral of the last term in \eqref{eq:SMfracaxion} an integer. Furthermore the term 
\begin{equation}
\tfrac{1}{72} \kappa_1 \int \mathcal{P}(\omega) \,,
\end{equation}
is well-defined modulo integers since $\tfrac{\kappa_1}{72} \in \tfrac{1}{12} \mathbb{Z}$ and $\int \mathcal{P}(\omega)$ is defined mod $12$. Integer instantons in the non-Abelian group factors, which are present for any value of $\mathcal{Z}$, further provide the conditions
\begin{equation}
\kappa_3 \,, \kappa_2 \in \mathbb{Z} \,.
\end{equation}
The remaining condition can be deduced by taking an integer lift of $\omega$ which we always assume to exist, see however footnote \ref{foot:cochain}, i.e., $\omega_{\mathbb{Z}} \in H^2 (M;\mathbb{Z})$ with
\begin{equation}
\int_C \omega_{\mathbb{Z}} \text{ mod } 6 = \int_C \omega \,, \quad \text{for all} \enspace C \in H_2 (M; \mathbb{Z}) \,,
\end{equation}
and notice that this is also a good integer lift of $\omega_i$
\begin{equation}
\int_C \omega_{\mathbb{Z}} \text{ mod } n = \int_C \omega \text{ mod } n = \int_C \omega_i \,, \quad \text{for all} \enspace C \in H_2 (M; \mathbb{Z}) \,.
\end{equation}
With this we find 
\begin{equation}
\big( \tfrac{1}{3} \kappa_3 + \tfrac{1}{4} \kappa_2 + \tfrac{1}{72} \kappa_1 \big) \int \omega_{\mathbb{Z}} \cup \omega_{\mathbb{Z}} \in \mathbb{Z} \,.
\end{equation}
Recalling that $\omega_{\mathbb{Z}} \cup \omega_{\mathbb{Z}}$ always integrates to an integer number on Spin manifolds the remaining quantization condition for the topological couplings $\kappa_i$ is given by
\begin{equation}
\tfrac{1}{3} \kappa_3 + \tfrac{1}{4} \kappa_2 + \tfrac{1}{72} \kappa_1 \in \tfrac{1}{2} \mathbb{Z} \quad \leftrightarrow \quad 24 \kappa_3 + 18 \kappa_2 + \kappa_1 \in 36 \mathbb{Z} \,,
\label{eq:smconstr}
\end{equation}
This precisely reproduces the result of \cite{Choi:2023pdp, Reece:2023iqn, Cordova:2023her}.

After the inclusion of $\mathfrak{g}_{\text{d}}$ this quantization can change, which we will demonstrate for $\mathcal{Z} = \mathbb{Z}_6$ and general dark sector gauge algebra. The embedding is defined by the parameters $(\ell_3 \,, \ell_2 \,, \ell_1 \,, \ell_{\text{d}})$. As above, integer instantons in all of the non-Abelian gauge groups give rise to the condition
\begin{equation}
\kappa_3 \,, \kappa_2 \,, \kappa_{\text{d}} \in \mathbb{Z} \,,
\end{equation}
leaving a single condition from fractional instantons. For cases with $\mathcal{Z}$ having several factors one gets a condition for each generator of $\mathcal{Z}$. It reads
\begin{equation}
\int \Big( \tfrac{\ell_3^2}{3} \kappa_3 \mathcal{P}(\omega_3) + \tfrac{\ell_2^2}{4} \kappa_2 \, \mathcal{P}(\omega_2) + \tfrac{\ell_1^2}{72} \kappa_1 \mathcal{P}(\omega) - \tfrac{\ell_1}{6} \kappa_1 \, \omega \cup c_1 (F_1) + \alpha_{\text{d}} \, \mathcal{P}(\omega_{\text{d}}) \Big) \in \mathbb{Z} \,.
\end{equation}
For this to possibly be an integer we need to demand that
\begin{equation}
\kappa_1 \in \tfrac{6}{\text{gcd}(\ell_1, 6)} \, \mathbb{Z} \,,
\end{equation}
which also ensures that the $\kappa_1 \mathcal{P}(\omega)$ term is well-defined modulo integers. Lifting $\omega$ and $\omega_i$ to $H^2(M; \mathbb{Z})$ the consistency condition is
\begin{equation}
\big( \tfrac{\ell_3^2}{3} \kappa_3 + \tfrac{\ell_2^2}{4} \kappa_2 + \tfrac{\ell_1^2}{72} \kappa_1 + \alpha_{\text{d}} \big) \int \omega_{\mathbb{Z}} \cup \omega_{\mathbb{Z}} \in \mathbb{Z} \,,
\end{equation}
which we express as
\begin{equation}
\tfrac{\ell_3^2}{3} \kappa_3 + \tfrac{\ell_2^2}{4} \kappa_2 + \tfrac{\ell_1^2}{72} \kappa_1 + \alpha_d \in \tfrac{1}{2} \mathbb{Z} \quad \leftrightarrow \quad 24 \ell_3^2 \, \kappa_3 + 18 \ell_2^2 \, \kappa_2 + \ell_1^2 \, \kappa_1 + 72 \alpha_{\text{d}} \in 36 \mathbb{Z} \,.
\label{eq:Z6embed}
\end{equation}
To go through the various possibilities, we provide a list of $\alpha_{\text{d}}$ for all $\mathfrak{g}_{\text{d}}$ and embeddings of $\mathbb{Z}_6$ specified by $\ell_{\text{d}}$ on Spin manifolds
\begin{equation}
\renewcommand*{\arraystretch}{1.5}
\begin{array}{| c | c | c | c |}
\hline
\text{gauge algebra } \mathfrak{g}_{\text{d}} & \text{gauged subgroup of } \mathcal{Z}_{\text{d}} & \ell_{\text{d}} & \alpha_{\text{d}} \\ \hline \hline
\mathfrak{so}(2n+1) & \mathbb{Z}_2 & 1 & \tfrac{1}{2} \kappa_{\text{d}} \\ \hline
\mathfrak{so}(4n) & \mathbb{Z}_2 & (1,0) & \tfrac{n}{4} \kappa_{\text{d}} \\ \hline
 & \mathbb{Z}_2 & (0,1) & \tfrac{n}{4} \kappa_{\text{d}}  \\ \hline
 & \mathbb{Z}_2 & (1,1) & \tfrac{1}{2} \kappa_{\text{d}}  \\ \hline
\mathfrak{so}(4n+2) & \mathbb{Z}_2 & 2 & \tfrac{1}{2} \kappa_{\text{d}}\\ \hline
\mathfrak{sp}(n) & \mathbb{Z}_2 & 1 & \tfrac{n}{4} \kappa_{\text{d}} \\ \hline
\mathfrak{e}_6 & \mathbb{Z}_3 & 1 \text{ or } 2 & \tfrac{2}{3} \kappa_{\text{d}} \\ \hline
\mathfrak{e}_7 & \mathbb{Z}_2 & 1 & \tfrac{3}{4}\kappa_{\text{d}} \\ \hline
\end{array}
\label{eq:fracdcoef}
\renewcommand*{\arraystretch}{1}
\end{equation}
Since $\alpha_{\text{d}}$ is only relevant modulo $\tfrac{1}{2}$ we find that the only non-trivial embeddings happen for $\mathfrak{so}(4n)$, $\mathfrak{sp}(n)$ for odd $n$ as well as $\mathfrak{e}_6$ and $\mathfrak{e}_7$. The most flexible embeddings happen for $\mathfrak{su}(n)$ dark sector gauge groups for which we find (up to $\mathfrak{su}(6)$)
\begin{equation}
\renewcommand*{\arraystretch}{1.5}
\begin{array}{| c | c | c | c |}
\hline
\text{gauge algebra } \mathfrak{g}_{\text{d}} & \text{gauged subgroup of } \mathcal{Z}_{\text{d}} & \ell_{\text{d}} & \alpha_{\text{d}} \\ \hline \hline
\mathfrak{su}(2) & \mathbb{Z}_2 & 1 & \tfrac{1}{4} \kappa_{\text{d}} \\ \hline 
\mathfrak{su}(3) & \mathbb{Z}_3 & 1 \text{ or } 2 & \tfrac{1}{3} \kappa_{\text{d}} \\ \hline
\mathfrak{su}(4) & \mathbb{Z}_2 & 2 & \tfrac{1}{2} \kappa_{\text{d}} \\ \hline
\mathfrak{su}(6) & \mathbb{Z}_2 & 3 & \tfrac{3}{4} \kappa_{\text{d}} \\ \hline
 & \mathbb{Z}_3 & 2 & \tfrac{2}{3} \kappa_{\text{d}} \\ \hline
 & \mathbb{Z}_6 & 1 \text{ or } 5 & \tfrac{5}{12} \kappa_{\text{d}} \\ \hline
\end{array}
\label{eq:fracdcoefsu}
\renewcommand*{\arraystretch}{1}
\end{equation}
One sees that the additional contribution of $\kappa_{\text{d}}$ generated by the topological mixing with the dark sector changes the quantization conditions.

In the next section we explore how this can influence the axion-photon interactions.

\subsection{Axion-photon coupling}
\label{subsec:axphot}

At energies below electro-weak symmetry breaking the weak SU$(2)$ and the hypercharge U$(1)_Y$ combine to the U$(1)$ of electro-magnetism, whose field strength we denote by $F$. The topological part of the action below this breaking scale is given by
\begin{equation}
\mathcal{L}'_{a,\text{top}} = i a \Big( \frac{N}{4 \pi^2} \, \text{tr} (F_3 \wedge F_3) + \frac{E}{8 \pi^2} \, F \wedge F + \frac{\kappa_{\text{d}}}{8 \pi^2} \, \text{tr} (F_{\text{d}} \wedge F_{\text{d}}) \Big) \,,
\end{equation}
where the constants $N$ and $E$ are often used in phenomenological discussions \cite{Srednicki:1985xd, DiLuzio:2020wdo} and are defined in terms of the $\kappa_i$ as
\begin{equation}
N = \tfrac{1}{2} \kappa_3 \,, \quad E = \tfrac{1}{36} (\kappa_1 + 18 \kappa_2) \,.
\end{equation}
From this one can also determine the axion-photon coupling which is given by ($\alpha$ is the fine-structure constant)
\begin{equation}
g_{a\gamma\gamma} = \frac{\alpha N}{\pi f} \Big( \frac{E}{N} - 1.92(4) + \rho \Big) \,.
\label{eq:axphot}
\end{equation}
Note that the numerical result $1.92$ arises from the mixing of the axion with the neutral pion, $\pi_0$, see the derivation in \cite{GrillidiCortona:2015jxo}. The particular realization of the dark sector might alter this numerical coefficient. However, since at leading order this mixing proceeds via Standard Model fields and we do not alter the electro-weak symmetry breaking, these effects will be suppressed with respect to the result stated above. Since further we do not want to include sensibility of the particular dark sector spectrum, we simply parameterize this modification with the constant $\rho$, which we expect to be sub-leading.

The important conclusion of \eqref{eq:axphot}, as pointed out in \cite{Reece:2023iqn, Agrawal:2023sbp}, is that there is a minimal axion-photon coupling, since one has to cancel the mixing contribution $\big(-1.92(4) + \rho\big)$ with a quantized $\tfrac{E}{N}$, which cannot be done with arbitrary precision. For example for the $\mathbb{Z}_6$ quotient in the Standard Model from \eqref{eq:smconstr} one has
\begin{equation}
\frac{E}{N} = \frac{\kappa_1 + 18 \kappa_2}{18 \kappa_3} = \frac{- 4 \kappa_3 + 6 m}{3 \kappa_3} \,, \quad \text{with } m \in \mathbb{Z} \,.
\end{equation}
Setting $\kappa_3 = 1$, which avoids potential domain wall problems, see Section \ref{sec:phenocons} below, the best cancellation with the pion mixing is achieved for $m = 2$, leading to $\tfrac{E}{N} = \tfrac{8}{3}$ and
\begin{equation}
g_{a\gamma\gamma}^{\text{min}} \simeq \frac{\alpha}{2 \pi f} \enspace 0.72(4) \,,
\end{equation}
and one obtains a lower bound of this coupling.

After the inclusion of the topological mixing with the dark sector, which we take to be induced by $\mathcal{Z} = \mathbb{Z}_6$ as above, one instead finds the relevant quantities to be
\begin{equation}
\frac{E}{N} = \frac{\kappa_1 + 18 \kappa_2}{18 \kappa_3} = \frac{6 m - 12 \alpha_{\text{d}} + 3(\ell_1^2 - \ell_2^2) \kappa_2 - 4 \ell_3^2 \kappa_3}{3 \ell_1^2 \kappa_3} \,, \quad \text{with} \enspace m \in \mathbb{Z} \,,
\end{equation}
where we used \eqref{eq:Z6embed} to eliminate $\kappa_1$. We already see that depending on the topological couplings $\kappa_i$ and the embedding parameters $\ell_i$ this equation can be tuned more finely and the minimal value for the axion-photon coupling can in turn be reduced significantly.

We will exemplify this for the example $\mathfrak{g} = \mathfrak{su}(3)$ with embedding given in \eqref{eq:exampleembedconf}, in which case $\alpha_{\text{d}} = \tfrac{\ell_{\text{d}}^2}{3} \kappa_{\text{d}} = \tfrac{1}{3} \kappa_{\text{d}}$ and one finds
\begin{equation}
\frac{E}{N} = \frac{6m - 4 \kappa_{\text{d}} - 4 \kappa_3}{3 \kappa_3} \,,
\end{equation}
again setting $\kappa_3 = 1$ one can obtain a significantly better cancellation in cases where $\kappa_{\text{d}} = - 1$ for which we can choose $m = 1$ to find
\begin{equation}
\frac{E}{N} = 2 \,,
\end{equation}
and the minimal value of the axion-photon coupling is given by
\begin{equation}
g_{a \gamma \gamma}^{\text{min,d}} \simeq \frac{\alpha}{2 \pi f} \enspace \big( 0.07(6) + \rho \big) \,.
\end{equation}
We see that while the direct dynamical consequences of a topological mixing with the dark sector are rather mild, it can have interesting consequences to the axion sector which is sensitive to these topological properties. In fact, assuming the corrections $\rho$ are sub-leading, the axion-photon coupling can be reduced by an order of magnitude in the presence of a dark $\mathfrak{su}(3)$ with non-trivial mixing with the Standard Model fields. Note further that even in situations where $\rho$ modifies the Standard Model calculations the additional parameter $\kappa_{\text{d}}$ leads to more freedom in tuning the parameters in order to achieve a better cancellation of the terms among each other.

\section{Symmetries and defects}
\label{sec:sym}

Systems with axionic degrees of freedom are very rich in terms of their symmetries. In particular they realize so-called higher-group structures, where various higher-form symmetries mix among each other \cite{Hidaka:2020iaz, Brennan:2020ehu, Hidaka:2020izy}, and non-invertible symmetries, which are symmetries that do not follow a group law \cite{Choi:2022jqy, Cordova:2022ieu, Choi:2022fgx, Yokokura:2022alv}. Some of these symmetries still exist after the coupling to the dark sector. Moreover, since the axion is a periodic scalar field there can be configurations with a non-trivial winding number, i.e., axion strings. Due to the topological coupling the axion string needs to host localized degrees of freedom accounting for the anomaly inflow from the bulk \cite{Callan:1984sa, Witten:1984eb}. For non-trivial coupling to the dark sector these localized degrees of freedom are also charged under the dark gauge group.

We will briefly discuss these properties of the symmetries and defects in the system in the following.

\subsection{The symmetries of the axion system}

In the absence of the topological coupling \eqref{eq:axcoupling} there are four different types of symmetries. The axion \eqref{eq:axkin} possesses a U$(1)$ 0-form shift symmetry
\begin{equation}
a \rightarrow a + \sigma \,, \text{ with } \sigma \in [0, 2 \pi) \,.
\label{eq:axionshift}
\end{equation}
and a winding U$(1)$ 2-form symmetry, which measures the winding number of the axion string configuration. 
The corresponding currents are
\begin{equation}\label{eq:symmetr}
    \ast j_{\text{shift}} = i f^2 \ast da \,, \quad  \ast j_{\text{wind}}= \tfrac{1}{2\pi}da
\end{equation}  
which are conserved, i.e., $d \ast j = 0$, prior to coupling of the axion to the gauge fields. In the modern language we would say that the shift symmetry is related to a 3-dimensional topological operator, which can be written in terms of the conserved current
\begin{equation}
U_{\sigma} (\Sigma_3) = e^{i \sigma \int_{\Sigma_3} \ast j_{\text{shift}}}
\end{equation}
where $\Sigma_3$ is the 3-dimensional submanifold of $M$, the topological operator is located on, see also \cite{Choi:2022jqy, Cordova:2022ieu, Choi:2022fgx, Yokokura:2022alv}.

On the other hand, as discussed in \ref{subsec:1gauging} the gauge theory sector contributes an electric center 1-form symmetry. It is the subgroup of 
\begin{equation}
\frac{\mathcal{Z}_{\widetilde{G}}}{\mathcal{Z}} = \frac{\mathbb{Z}_3 \times \mathbb{Z}_2 \times \text{U}(1) \times \mathbb{Z}_{\text{d}}}{\mathcal{Z}} \,,
\end{equation}
which is not broken by the presence of dynamical charged matter states, as for example provided by the Standard Model matter fields. On the other hand the gauging process introduces a dual magnetic 1-form symmetry $\mathcal{Z}^{\vee}$, the Poincar\'e dual of $\mathcal{Z}$, which acts non-trivially on 't Hooft lines that cannot be screened by dynamical magnetic monopoles \cite{Aharony:2013hda}.

In the presence of the interaction \eqref{eq:axcoupling} the equations of motion are modified and the current for the shift symmetry is not conserved anymore, instead one has
 \begin{equation}
d\ast j_{\text{shift}} = \frac{\kappa_3}{8\pi^2}\text{tr}(F_3\wedge F_3) + \frac{\kappa_2}{8\pi^2}\text{tr}(F_2\wedge F_2) + \frac{\kappa_1}{8\pi^2} F_1\wedge F_1 + \frac{\kappa_\text{d}}{8\pi^2}\text{tr}(F_\text{d}\wedge F_\text{d}) \,,
\label{eq:symmbreak}
\end{equation}
and the continuous symmetry is broken. However, it might not be broken completely. To see that, consider a shift \eqref{eq:axionshift} and define the modified operator
\begin{equation}
\widetilde{U}_{\sigma} (\Sigma_3) = e^{i \sigma \int_{\Sigma_3} (\ast j_{\text{shift}} - \frac{\kappa_3}{2\pi} \Omega_3-\frac{\kappa_2}{2\pi} \Omega_2 - \frac{\kappa_1}{8\pi^2}A_1\wedge F_1 - \frac{\kappa_d}{2\pi} \Omega_d)} \,,
\label{eq:modshiftop}
\end{equation}
with non-Abelian Chern-Simons 3-forms $\Omega_i = \frac{1}{4\pi}\text{tr}(A_i\wedge dA_i+\frac{2}{3}A_i^3)$, which satisfy
\begin{equation}
d\Omega_i=\tfrac{1}{4\pi}\text{tr}(F_i\wedge F_i) \,.
\end{equation}
Since the Chern-Simons terms are not invariant under large gauge transformations the terms $\sigma \, \kappa_i \Omega_i$ are in general not well-defined, even though the resulting operator in \eqref{eq:modshiftop} is topological. An alternative way to define \eqref{eq:modshiftop} is to define a 4-dimensional submanifold $\Sigma_4$ of $M$ with boundary $\partial \Sigma_4 = \Sigma_3$ and write
\begin{equation}
\widetilde{U}_{\sigma} (\Sigma_3) = e^{i \sigma \big( \int_{\Sigma_3} \ast j_{\text{shift}} - \int_{\Sigma_4} ( \frac{\kappa_3}{8\pi^2}\text{tr}(F_3\wedge F_3) + \frac{\kappa_2}{8\pi^2}\text{tr}(F_2\wedge F_2) + \frac{\kappa_1}{8\pi^2} F_1 \wedge F_1 + \frac{\kappa_\text{d}}{8\pi^2}\text{tr}(F_\text{d}\wedge F_\text{d})) \big)} \,.
\label{eq:extendeddef}
\end{equation}
For that to be a well-defined 3-dimensional topological operator, we need to demand that $\widetilde{U}_{\sigma}$ does not depend on $\Sigma_4$, which means that it should be trivial whenever $\Sigma_4$ does not have a boundary. The special values for $\sigma$ for which this is the case, for given $\kappa_i$ defines the unbroken part of the U$(1)$ shift symmetry. The requirement reads
\begin{equation}
     \sigma \, \int_{\Sigma_4} \Big( \tfrac{\kappa_3}{8\pi^2}\text{tr}(F_3\wedge F_3)+\tfrac{\kappa_2}{8\pi^2}\text{tr}(F_2\wedge F_2)+\tfrac{\kappa_1}{8\pi^2}F_1\wedge F_1+\tfrac{\kappa_\text{d}}{8\pi^2}\text{tr}(F_\text{d}\wedge F_\text{d})\Big)\in 2\pi \mathbb{Z} \,.
\label{eq:fracsymm}
\end{equation}
This precisely corresponds to the discussion of fractional instantons in Section \ref{subsec:quant} and the quantization condition in \eqref{eq:axionquant}. As we saw there in the presence of the nontrivial 1-from background $\omega \in  H^2(M; \mathcal{Z})$ we can rewrite condition \eqref{eq:fracsymm}
\begin{align}\label{eq:fracsymm2}
\begin{split}
    \sigma \Big(&\kappa_3 n_3^{\text{int}}++\kappa_2 n_2^{\text{int}} + \kappa_1 n_1^{\text{int}} + \kappa_{\text{d}} n_{\text{d}}^{\text{int}}+ \\
    & \int_{\Sigma_4} \big(\alpha_3 \, \mathcal{P} (\omega_3) + \alpha_2 \, \mathcal{P}(\omega_2) +  \alpha_1 \, \mathcal{P}(\omega) - \widetilde{\alpha}_1 \, \omega \cup c_1(F_1) + \alpha_{\text{d}} \, \mathcal{P}(\omega_{\text{d}}) \big)\Big) \in 2\pi \mathbb{Z} \,,
\end{split}
\end{align}
where we have split the expression into the integral instanton contributions $n_i^{\text{int}}$ and the fractional contributions with $\alpha_i$ defined as in \eqref{eq:axfrac}. From the consistency of the axion we know that the second line has to be an integer. There is an unbroken shift symmetry in case this integer as well as the prefactors $\kappa_3$, $\kappa_2$, $\kappa_1/6$, and $\kappa_{\text{d}}$ are divisible by a common integer. The remaining shift symmetry is given by $\mathbb{Z}_p$ with shifts $\sigma = \tfrac{2\pi}{p}$ and $p$ in the case of $\mathcal{Z} = \mathbb{Z}_6$, defined in \eqref{eq:Z6embed}, given by
\begin{equation}
p = \text{gcd} \Big( \kappa_3 \,, \kappa_2 \,, \frac{\kappa_1}{6} \,, \kappa_{\text{d}} \,, \frac{24 \ell_3^2 \kappa_3 + 18 \ell_2^2 \kappa_2 + \ell_1^2 \kappa_1 + 72 \alpha_{\text{d}}}{36} \Big) \,.
\end{equation}
As discussed in \cite{Choi:2022jqy, Cordova:2022ieu, Choi:2022fgx, Yokokura:2022alv}, for other special values of shifts $\sigma$ it is possible to resurrect part of the broken shift symmetries as non-invertible symmetries. For that one compensates the non-invariance under gauge transformation of the topological operators in \eqref{eq:modshiftop} by including a topological field theory sector, that couples to the background gauge fields to make the whole system gauge invariant. Consider a U$(1)$ gauge field $c$ living on the topological operators localized on $\Sigma_3$ with the topological level $k$ Chern-Simons action
\begin{equation}\label{eq:cscor}
    S=\frac{ik}{4\pi}\int_{\Sigma_3} \big( c\wedge dc +2 c \wedge b \big) \,,
\end{equation}
where $b = \tfrac{2 \pi}{k} \omega$ is the U$(1)$ realization of $\omega \in H^2 (M;\mathbb{Z}_k)$ as discussed in \eqref{eq:disccont}. Since $c$ appears only quadratically in the effective action \eqref{eq:cscor}, one can integrate it out using its equation of motion $dc+b=0$. Substituting this back in the action and using the extension to $\Sigma_4$ as in \eqref{eq:extendeddef} one finds
\begin{equation}
S= -\frac{ik}{4\pi} \int_{\Sigma_4} b\wedge b= -\frac{2\pi i }{2k}\int_{\Sigma_4}\mathcal{P}(\omega) 
\end{equation}
compensating precisely for a fractional instanton in one of the gauge sectors. In our case this allows the discussion of a non-invertible shift symmetry with shifts $\sigma = \tfrac{2 \pi}{q}$, where now
\begin{equation}
q = \text{gcd} (\kappa_3 \,, \kappa_2 \,, \kappa_{\text{d}}) \,.
\end{equation}
Note that the coefficient $\kappa_1$ for the Abelian gauge factor does not show up any more, since for U$(1)$ one can recover similarly, utilizing \eqref{eq:cscor} as above, the full shift symmetry, or at least its rational part $\mathbb{Q}/\mathbb{Z}$, see \cite{Choi:2022jqy, Cordova:2022ieu}. Since the cancellation of the fractional instanton contribution happens separately for each gauge sector, we need Chern-Simons terms of the form above for each gauge group factor, including $\widetilde{G}_{\text{d}}$ that allows for fractional instanton contributions, whose gauge fields $c_i$ couple to the non-trivial background $\omega_i$, respectively. In general we notice that in the presence of topological mixing with a dark sector the unbroken invertible and non-invertible shift symmetries are smaller than what would be inferred from the Standard Model couplings alone.

The topological coupling of the axion further modifies the electric 1-form symmetries, as discussed in \cite{Choi:2022fgx, Yokokura:2022alv}, since it modifies the equation of motion for the gauge field as well. The specific form of the (non-invertible) electric 1-form symmetries, however, depends on the spectrum of dynamical particles, including the exotic matter representations discussed in Section \ref{subsec:reps} and is therefore model dependent. For a detailed account of this in the context of the Standard Model we refer to \cite{Choi:2023pdp}.

The 0-form, 1-form, and 2-form symmetries of the system mix into a higher group structure as was analyzed in \cite{Hidaka:2020iaz, Brennan:2020ehu, Hidaka:2020izy}. Once more, the specific realization of this depends on the matter representations of the system and needs to be analyzed on a case by case basis. For the Standard Model the result was derived in  \cite{Choi:2023pdp}. This higher group structure is also reflected in the anomaly inflow on various defects of the axion system which we will briefly discuss in the following. 
 
\subsection{The defects of the axion system}

Theories with axions have extended objects. Among them are codimension-one objects, i.e., domain walls, across which the value of the axion field jumps, and codimension-two objects, axion strings, around which the axion value winds $a \rightarrow a + 2\pi$. Comparing this behavior of the axion field with the definition of the generalized symmetries in \eqref{eq:symmetr} we see that dynamical axion strings break the 2-form winding symmetry.

As discussed above, if $q = \gcd(\kappa_2, \kappa_3, \kappa_{\text{d}})>1$, there is an exact $\mathbb{Z}_{q}$ symmetry which, even though it could be non-invertible, still acts on axions as the standard shift symmetry $a \rightarrow a + \frac{2\pi}{q}$. This implies that axion potential has to be $\tfrac{2 \pi}{q}$-periodic. Another way to argue for this $\tfrac{2 \pi}{q}$-periodicity of the potential is by considering instanton contribution to the computation of the axion potential. Since it is expected that the axion potential comes from summing over all instantons $e^{i\kappa_i n_i a}$ for integral instanton numbers $n_i$. The Abelian and fractional instantons do not contribute to the axion potential for Minkowski spacetime \cite{Cordova:2023her}, since they need non-trivial spacetime topology.\footnote{This changes once one includes topology changes that are expected to occur for quantum gravity.} Thus, below the electro-weak phase transition the only contributions come from the strong interactions, SU$(3)$, and potentially the dark sector $\widetilde{G}_{\text{d}}$ in case it confines. If it does, one expects that there will be contributions of order $\Lambda_{\text{d}}^4$ to the potential and the potential has approximate $\tfrac{2 \pi}{\text{gcd}(\kappa_3,\kappa_{\text{d}})}$-periodicity imposed by the non-invertible shift symmetry. Therefore there are degenerate vacua and domain walls in case $\text{gcd}(\kappa_3, \kappa_{\text{d}}) > 1$. If, on the other hand, the dark sector gauge group is broken the axion potential is approximately  $\tfrac{2\pi}{\kappa_3}$-periodic and one obtains $|\kappa_3|$ vacua between the axion domain walls interpolate. In both cases the vacuum expectation value of the axion spontaneously breaks the non-invertible shift symmetry \cite{Cordova:2023her}.

This further demonstrates the close relation between axion domain walls and the symmetry operators for the shift symmetry. As we have seen, in case the shift symmetry is non-invertible this requires the existence of a topological field theory sector on the worldvolume of the domain wall. These topological field theories, related to the fractional quantum Hall effect, contain anyons which feel the influence of the background fields of both the Standard Model and dark sector 1-form symmetries. This can also be interpreted in terms of an anomaly inflow of the 1-form symmetries, see, e.g., \cite{Dierigl:2014xta, Gaiotto:2014kfa, Gaiotto:2017yup}.

There is also anomaly inflow, this time of the 0-form gauge symmetries, in the presence of axion strings \cite{Callan:1984sa, Witten:1984eb}, see also \cite{Brennan:2023kpw} for a recent discussion. This implies that the string hosts local degrees of freedom that cancel this contribution. To see this inflow explicitly we can perform a 0-form gauge transformation and 1-form transformations in the presence of an axion string localized on a 2-dimensional manifold $\Sigma_2$. This means that the value of the axion field on paths winding around this surface undergoes the monodromy $a \rightarrow a + 2 \pi$. Locally, one can write that as
\begin{equation}
d^2a= 2 \pi \, \delta_{\Sigma_2} \,,
\label{eq:axionstring}
\end{equation}
where $\delta_{\Sigma_2}$ is a 2-form which is localized on $\Sigma_2$, i.e., the Poincar\'e dual of $\Sigma_2$, for a more explicit treatment with bump functions see \cite{Harvey:2000yg}. Considering the topological interaction of the axion \eqref{eq:axcoupling}, we find locally
\begin{equation}
\mathcal{L}^a_{\text{top}} = - i \, da \wedge \big( \tfrac{\kappa_3}{2 \pi} \Omega_3 + \tfrac{\kappa_2}{2 \pi} \Omega_2 + \tfrac{\kappa_1}{8 \pi^2} A_1 \wedge F_1 + \tfrac{\kappa_{\text{d}}}{2 \pi} \Omega_{\text{d}} \big) \,.
\end{equation}
Under infinitesimal gauge transformations $\lambda_i$ the action transforms as
\begin{equation}
\Delta S = - \int_{M} \tfrac{i}{8\pi^2} \, da \wedge \big( \kappa_3 \, \text{tr}(d \lambda_3 \wedge F_3) + \kappa_2 \, \text{tr}(d \lambda_2 \wedge F_2) + \kappa_1 \, d \lambda_1 \wedge F_1 + \kappa_{\text{d}} \, \text{tr}(d \lambda_{\text{d}} \wedge F_{\text{d}}) \big) \,.
\end{equation}
Integrating this expression by parts and using \eqref{eq:axionstring} one finds a gauge variation localized to the worldvolume of the axion string $\Sigma_2$
\begin{equation}
\Delta S = - \int_{\Sigma_2} \tfrac{i}{4 \pi} \big( \kappa_3 \, \text{tr}(\lambda_3 F_3) + \kappa_2 \, \text{tr}(\lambda_2 F_2) \big) + \kappa_1 \, \lambda_1 F_1 + \kappa_{\text{d}} \, \text{tr}(\lambda_{\text{d}} F_{\text{d}}) \big) \,,
\end{equation}
due to anomaly inflow. This needs to be cancelled by the existence of charged chiral fields on $\Sigma_2$. Assuming the existence of two-dimensional chiral fermions $\psi$ transforming in the representation $\mathbf{R} = (\mathbf{R}_3 \,, \mathbf{R}_2 \,, \mathbf{R}_{\text{d}})_q$ of $\widetilde{G}$ they contribute to the gauge variation due to their perturbative anomaly as 
\begin{align}
\begin{split}
\Delta S_f =& \int_{\Sigma_2} \sum_{\mathbf{R}} \frac{i}{4\pi}\Big( \text{dim}(\mathbf{R}_2,\mathbf{R}_{\text{d}}) I(\mathbf{R}_3) \, \text{tr} (\lambda_3 F_3) + \text{dim}(\mathbf{R}_3,\mathbf{R}_{\text{d}}) I(\mathbf{R}_2) \, \text{tr}(\lambda_2 F_2 ) + \\
& \hspace{1.55cm} \text{dim}(\mathbf{R}_3,\mathbf{R}_2) I(\mathbf{R}_{\text{d}}) \, \text{tr}(\lambda_{\text{d}} F_{\text{d}}) + \text{dim}(\mathbf{R}_3, \mathbf{R}_2, \mathbf{R}_{\text{d}}) 36 q^2 \, \lambda_1 F_1 \Big) \,,
\end{split}
\end{align}
where $I(\mathbf{R})$ is the Dynkin index such that $\text{tr} (T^a_{\mathbf{R}} T^b_{\mathbf{R}}) = \tfrac{1}{2} I(\mathbf{R}) \delta^{ab}$ and the sum runs over representations of all chiral fermions on $\Sigma_2$. Note that these localized fermions can also arise as localized zero modes of Dirac fermions in four dimensions and therefore might be sensitive to even heavy exotic matter particles, see also \cite{Witten:1984eb}. Demanding that
\begin{equation}
\Delta S + \Delta S_f = 0 \,,
\end{equation}
one finds the consistency conditions
\begin{equation}
\sum_{\mathbf{R}} \frac{\text{dim}(\mathbf{R}_3, \mathbf{R}_2, \mathbf{R}_{\text{d}})}{\text{dim}(\mathbf{R}_i)} I(\mathbf{R}_i) = \kappa_i \,, \quad 36 \sum_{\mathbf{R}} \text{dim}(\mathbf{R}_3, \mathbf{R}_2, \mathbf{R}_{\text{d}}) q^2 = \kappa_1 \,,
\end{equation}
for non-Abelian and Abelian variations, respectively. Therefore, we see that if $\kappa_{\text{d}}$ does not vanish the axionic strings necessarily host chiral degrees of freedom that are charged under the dark sector.

Similarly, we can perform a 1-form gauge transformation in the topological action. Let us perform this analysis explicitly with the example of a single gauge factor SU$(n)$ on which $\mathcal{Z} = \mathbb{Z}_k$ acts non-trivially. Expressing the $H^2(M;\mathcal{Z})$ gauge field as continuous $b$, see \eqref{eq:disccont}, one can rewrite the topological coupling using \eqref{eq:nAinstantonmod} as
\begin{equation}
\mathcal{L}^a_{\text{top}} \supset  \tfrac{i \kappa}{8 \pi^2} a \,
\text{tr}((F'-b\mathbf{1}) \wedge (F'-b \mathbf{1}))= \tfrac{i \kappa}{8 \pi^2} a \big( 
\text{tr}(F' \wedge F')-n b\wedge b \big)\,,
\end{equation}
where the $F'$ is a $U(n)$ gauge field as in \eqref{eq:ungauge} . Moreover, locally we can express the closed 2-form field $b$ in terms of a U$(1)$ 1-form gauge field $c$ as
\begin{equation}
k b = d c \,.
\end{equation}
The 1-form gauge symmetry transformations are then implemented by
\begin{equation}
A' \rightarrow A' + \Lambda \mathbf{1}  \,, \quad b \rightarrow b + d \Lambda \,, \quad c \rightarrow c + k  \Lambda \,,
\end{equation}
with 1-form gauge parameter $\Lambda$, see \cite{ Gaiotto:2014kfa, Gaiotto:2017yup }. After expressing $b$ in terms of $c$ we can aim to use the same trick as for the 0-form symmetries above, i.e., integrating by parts then doing the gauge variation and integrating by part once more to find the localized anomaly. However, we find that after integration by parts
\begin{equation}\label{eq:anomvan}
S \supset - \frac{i \kappa}{2\pi} \int da \wedge \big( \Omega - \tfrac{n}{4 \pi k^2} \, c \wedge dc \big)
\end{equation}
the action is invariant under 1-form gauge transformations and there is no inflow of an anomaly for the 1-form symmetries at the local level. Analogously, with a product of multiple simple gauge groups like in our setup, one has a cancellation for each separate gauge sector, analogous to \eqref{eq:anomvan}. 

Once more this analysis shows what a diverse system, in terms of its symmetries, anomalies, and defects, the axion model is.

\section{Phenomenological consequences}
\label{sec:phenocons}

The fact that the axion now couples to more than the Standard Model gauge groups also has consequences on the phenomenological implications, which we want to mention in the following. This is particularly important for the confining scenario, since it leads to major modifications to the axion potential of order $\Delta V_{a} \sim \Lambda_{\text{d}}^4$, which is the model-dependent dynamical scale of the dark sector. In the scenario of a broken dark sector gauge group all corrections are suppressed by the symmetry breaking scale $\Gamma_{\text{d}}$ and therefore can be made small.

\subsection{Solution to the strong CP problem}

Famously, the QCD axion offers a dynamical solution to the strong CP problem \cite{Weinberg:1977ma, Peccei:1977hh, Wilczek:1977pj, Peccei:1977ur}, i.e., its vacuum expectation value minimizes the CP-violation generated by a bare QCD $\theta$-angle, whose effective value is experimentally constrained to be
\begin{equation}
\overline{\theta} \lesssim 10^{-10} \,.
\end{equation}
This works if the axion potential is generated by the strong QCD dynamics but can be modified after the inclusion of further interactions, this is also known as the axion quality problem, see, e.g., \cite{Reece:2023czb}.

For a broken dark sector gauge group the corrections are suppressed by $\Gamma_{\text{d}}$ and it depends on the details of the setup, whether the corrections are small enough to allow for a dynamical solution for the strong CP problem. For a confining dark sector the situation is significantly worse, since there will be corrections to the axion potential of order $\Lambda_\text{d}^{4}$, for $\Lambda_\text{d}$ a strong coupling scale of the dark sector,  which will generically shift the minimum of the potential away from $\overline{\theta} = 0$. The only scenario for which this does not seem to happen is if the contributions to the axion potential from both sectors align in such a way that the minimum remains at the CP preserving value for QCD. For example this would require an alignment in the bare theta angles $\theta$ and $\theta_{\text{d}}$ to a very high precision, which seems not to be natural without additional symmetries demanding that.

\subsection{Axion domain wall problem}

As discussed above, if the axion only couples to the Standard Model, there are stable axion domain walls for $\kappa_3 > 1$. This is due to the fact that the axion potential has periodicity $\frac{2 \pi}{\kappa_3}$ and thus there are $|\kappa_3|$ distinct minima. Axion profiles that interpolate between two of such minima form stable domain walls.\footnote{These domain walls become unstable once the total shift in the axion value is a multiple of $2\pi$ for which they can form holes bounded by axion strings.} The energy density of a single domain wall is so high that it would lead to inconsistencies in the cosmological evolution of our universe, which is called the axion domain wall problem \cite{Sikivie:1982qv}, which also can be phrased in terms of domain wall networks see, e.g., \cite{Reece:2023czb, Cordova:2023her}. A solution to this problems is to set $\kappa_3 = 1$, which we have also done for the analysis in Section \ref{subsec:axphot}, in which case there are no stable domain walls. Alternatively, one can dilute the density of these objects by mechanisms like inflation, if they form early enough.

Once there is a confining dark sector there are important further contributions to the axion potential and one finds various parameter regimes according to the hierarchy between the two confinement scales $\Lambda_{\text{QCD}}$ and $\Lambda_{\text{d}}$:
\begin{itemize}
    \item{$\Lambda_{\text{QCD}} \gg \Lambda_{\text{d}}$: In this regime the axion potential is dominated by the dynamics of the strong force. This also applies for the number of (approximate) minima and we expect to have a potential domain wall problem for $\kappa_3 > 1$. Nevertheless the dark sector contributions to the potential generically lift the individual minima, which might extend the allowed parameter space for phenomenologically viable models.}
    \item{$\Lambda_{\text{d}} \gg \Lambda_{\text{QCD}}$: The axion potential is dominated by the dark sector potential and therefore $\kappa_3 > 1$ does not necessarily lead to a domain wall problem. However, there can be a domain wall problem in case $\kappa_{\text{d}} > 1$.}
    \item{$\Lambda_{\text{QCD}} \sim \Lambda_{\text{d}}$: In this regime the form of the potential depends on many of the details, such as the bare $\theta$ angles of the system, and one needs to analyse the specific models.}
\end{itemize}
We see that the presence of a dark sector can weaken the axion domain wall problem and open up new parameter regimes. It even offers a solution to the axion domain wall problem without requiring $\kappa_3 = 1$.

\subsection{Dark matter}

Axions also have appeared as viable dark matter candidate \cite{Preskill:1982cy, Dine:1982ah, Abbott:1982af}, but it is not clear what the fraction of dark matter formed by axions is. This contribution of axion dark matter will be highly sensitive to the specific scenario and therefore requires a model dependent analysis, however, the presence of a dark sector typically opens up new interesting regions of parameter space. Moreover, in the case of a confining dark sector one further has natural other candidates for dark matter at mass scales set by $\Lambda_{\text{d}}$, which can correspond to dark baryons, or dark glueballs in the absence of matter states, \cite{Strassler:2006im, Bai:2013xga, Boddy:2014yra}. Of course for that it is important that the confined dark objects making up dark matter are neutral under the Standard Model gauge group and are produced appropriately.

\vspace{0.5cm}

Summarizing, the presence of the dark sector poses a challenge for the dynamical solution of the strong CP problem, but might open up new phenomenologically interesting regimes for viable model building. 

\section{Conclusions}
\label{sec:concl}

In this manuscript we explored several effects of having a non-Abelian dark sector gauge group mixing topologically with the Standard Model fields. The specification of the global form of the gauge group imposes restrictions on the allowed matter representations and therefore influences the spectrum of potential exotic matter particles. Moreover, the topological mixing extends the allowed gauge field configurations which need to be included in the partition function, these include the appearance of fractional instantons that influence the dynamics of axion fields that couple to the instanton density. The periodicity of the axion leads to quantization conditions of these couplings which in turn modifies the possible values of the axion-photon coupling. In this way the presence of a dark sector can lead to a significant reduction of such a coupling compared to the Standard Model analysis, which opens up a larger parameter space. Finally, the topological mixing further influences the symmetries and defects of the axion system which further influences its phenomenological implications.

The setup we study allows for various generalizations. For example one could include several dark sector gauge groups that mix topologically with the Standard Model. This would further modify the quantization conditions and can even further reduce the axion-photon coupling. Nevertheless, since the fractional instanton numbers are rather constrained, we expect that mixing with a single gauge group already captures most of the allowed reduction. Of course the freedom to let part of the dark sector confine and break another part further increases the flexibility in phenomenologically interesting scenarios.

A more drastic modification is the inclusion of more than one axion field, which is well motivated from a string theory perspective, see, e.g., \cite{Arvanitaki:2009fg}. In this case the periodicity condition for each of the axion fields will lead to quantization conditions for its topological coupling. If the dark sector confines at a high scale, one can integrate out the heavy axion combination coupling to its instanton density, with the other light linear combinations only coupling to the Standard Model gauge fields. Since $\kappa_{\text{d}}$ for the remaining axion vanishes, they are only sensitive to the topological mixing within the Standard Model gauge group, consistent with decoupling.\footnote{We thank Matt Reece for emphasizing this aspect.} Moreover, the now larger dimensional field space for the axion potential allows for a richer structure of minima and might save axions as a dynamical solution to the strong CP problem. Thus, it would be interesting to see the effect of topological mixing, which is very common at least in a large number of dimensions \cite{Cvetic:2020kuw, Font:2020rsk, Font:2021uyw, Cvetic:2022uuu, Gould:2023wgl}, see also \cite{Cvetic:2023pgm, Baume:2023kkf}, in explicit string theory constructions with axions, which also allows for a statistical analysis \cite{Demirtas:2021gsq, Gendler:2023kjt}. This might also shed some light on the fate of the rich variety of generalized symmetries of axion systems in the context of a quantum gravity theory.

While in this work we focused on the general properties and implications of topological mixing with a dark sector, the next step would be to analyze promising specific models in more detail. This includes an estimation of the dark matter abundance in a confined dark sector scenario \cite{Strassler:2006im, Bai:2013xga, Boddy:2014yra} as well as estimates for the stability of axion domain walls \cite{Cordova:2023her}. Also a hierarchy of breaking scales, now involving the characteristic energy scales of the dark sector, along the lines of \cite{Cordova:2018cvg, Brennan:2020ehu, Choi:2022fgx, Choi:2023pdp} offers the possibility of constraints on allowed generalized symmetries for axion systems with viable phenomenology.

Throughout our analysis we have restricted to spacetime manifolds that allow for a Spin structure. While for the Standard model this seems to be the natural choice, one might ask which conclusions change in more general setup. In particular, theories for which spacetime does not have to be orientable, see \cite{McNamara:2022lrw}, or in which the Spin structure is replaced by a more general tangential structure that allows for (charged) fermions \cite{Putrov:2023jqi} are natural places to look for extensions of the models with Spin structure considered here.

Finally, it would be very interesting to explore a top-down explanation of when the bare $\theta$-angles of different gauge sectors have to be aligned, or whether there exist mechanisms to guarantee such an alignment. Such a mechanism might also point towards alternative solutions to the strong CP problem which does not need the introduction of dynamical matter fields, see, e.g., \cite{Cecotti:2018ufg}.

\section*{Acknowledgements}

MD thanks the ESI in Vienna, in particular the program ``The Landscape vs. the Swampland'', for hosting him during part of the time in which this work was completed. MD would like to thank Jakob Moritz for fruitful discussions. We are especially grateful to Matt Reece, for very valuable comments on the draft.

\begin{appendix}

\section{Fractional instantons for embedding of subgroups}
\label{app:fracinstsub}

In this appendix we investigate the fractional instanton number in situations in which the map from the quotient group $\mathcal{Z}$ into the center of the gauge group factor $\widetilde{G}_i$ only produces a subgroup of $\mathcal{Z}_{\widetilde{G}_i}$.

If $\widetilde{G}_i$ is non-Abelian the only relevant groups for which $\mathcal{Z}_{\widetilde{G}_i}$ admits non-trivial subgroups are, see Table \eqref{eq:groups},
\begin{equation}
\mathcal{Z}_{\text{SU}(n)} = \mathbb{Z}_n \,, \quad \mathcal{Z}_{\text{Spin}(4n)} = \mathbb{Z}_2 \times \mathbb{Z}_2 \,, \quad \mathcal{Z}_{\text{Spin}(4n+2)} = \mathbb{Z}_4 \,,
\end{equation}
for which we analyse the fractional instanton numbers individually.

For $\widetilde{G}_i = \text{Spin}(4n)$ there are three different $\mathbb{Z}_2$ subgroups of the center generated by
\begin{equation}
(1,0) \,, \quad (0,1) \,, \quad (1,1) 
\end{equation}
as elements in $\mathbb{Z}_2 \times \mathbb{Z}_2$. Identifying the background fields of the two $\mathbb{Z}_2$ factors with $B_L$ and $B_R$, and introducing $\omega \in H^2 (M;\mathbb{Z}_2)$ as a summation index in the partition function, we can use Table \eqref{eq:fracinst} to find
\begin{align}
\begin{split}
(1,0):& \quad n^{\text{frac}} = \tfrac{n}{4} \int \mathcal{P}(\omega) \text{ mod } 1 \,, \\
(0,1):& \quad n^{\text{frac}} = \tfrac{n}{4} \int \mathcal{P} (\omega) \text{ mod } 1 \,,
\end{split}
\end{align}
which on Spin manifolds can only be fractional, with non-trivial values $\tfrac{1}{2}$, for $n \in \{ 1 , 3\}$ mod 4. For $n = 1$ this has a simple explanation, since Spin$(4) \cong \text{SU}(2) \times \text{SU}(2)$ and the $\mathbb{Z}_2$ factors appear as center symmetries for the SU$(2)$'s. For the last embedding we find
\begin{equation}
(1,1): \quad n^{\text{frac}} = \tfrac{1}{2} \int \omega \cup \omega \,,
\end{equation}
which is not fractional on Spin manifolds.

For $\widetilde{G}_i = \text{Spin}(4n+2)$ one has non-trivial subgroup $\mathbb{Z}_2$ and the summation is given by the embedding
\begin{equation}
B = 2 \omega \,,
\end{equation}
where $B \in H^2(M;\mathbb{Z}_4)$ and $\omega \in H^2(M;\mathbb{Z}_2)$. Assuming $\omega$ has a lift to an element in $H^2(M;\mathbb{Z})$, denoted by $\omega_{\mathbb{Z}}$ one has
\begin{equation}
\int \mathcal{P}(2 \omega) = 4 \int \omega_{\mathbb{Z}} \cup \omega_{\mathbb{Z}} \text{ mod } 8 \,,
\end{equation}
which vanishes on Spin manifolds. So there are no fractional instanton numbers for gauge group Spin$(4n+1)/\mathbb{Z}_2$.

Finally, we consider $\widetilde{G}_i = \text{SU}(n)$. For that to have non-trivial subgroup $n$ cannot be prime and we want to consider fractional instanton numbers for gauge group SU$(n)/\mathbb{Z}_k$. This happens in case the embedding parameter $\ell_i$ satisfies
\begin{equation}
\text{gcd}(\ell_i, n) \neq 0 \,,
\end{equation}
and we can identify
\begin{equation}
k = \frac{n}{\text{gcd}(\ell_i,n)} \,.
\end{equation}
We can define the 1-form symmetry background via an element $\ell_i \, \omega \in H^2(M; \mathbb{Z}_n)$ with $\omega \in H^2(M; \mathbb{Z}_k)$.
 With $\ell_i \, \omega$ we can use the usual formula as indicated in Table \eqref{eq:fracinst} 
\begin{equation}
n^{\text{frac}} = \frac{n-1}{2n} \int \mathcal{P} (\ell_i \, \omega) = \ell_i^2 \, \frac{n-1}{2n} \int \mathcal{P} (\omega) = \frac{\ell_i^2}{\big(\text{gcd}(\ell_i,n)\big)^2} \, \frac{n (n-1)}{2 k^2} \int \mathcal{P} (\omega) \,,
\end{equation}
for which the fractional parts match. The first $\mathcal{P}$ is considered with respect to a $\mathbb{Z}_n$ class, while in the latter two with respect to $\mathbb{Z}_n$ class.\footnote{One should lift $\ell_i \omega$ to its integral representative and use co-chains instead of co-cycles to make sense of the transition \cite{Kapustin:2013qsa}.}

For Abelian $\widetilde{G}_i = \text{U}(1)$ gauge groups for which the center is the group itself, it is clear that the image of $\mathcal{Z}$ is just a subgroup defined by
\begin{equation}
\mathcal{Z} \supset \mathbb{Z}_n: \quad 1 \text{ mod } n \mapsto e^{2 \pi i \frac{\ell_i}{n}} \,.
\end{equation}
The calculation follows that around \eqref{eq:Abinstanton} and shows that the fractional part of the Abelian instanton is given by
\begin{equation}
n^{\text{frac}} = \int \big( \tfrac{\ell}{n} c_1(F) \cup \omega + \tfrac{\ell^2}{n^2} \mathcal{P}(\omega) \big) \,.
\end{equation}

\end{appendix}

\bibliographystyle{JHEP.bst}
\bibliography{papers.bib}

\end{document}